\documentclass[fleqn,usenatbib,onecolumn]{mnras}

\usepackage[T1]{fontenc}

\DeclareRobustCommand{\VAN}[3]{#2}
\let\VANthebibliography\thebibliography
\def\thebibliography{\DeclareRobustCommand{\VAN}[3]{##3}\VANthebibliography}

\usepackage{graphicx}	
\usepackage{amsmath}	
\usepackage{amssymb}	
\usepackage{bm}         
\usepackage{mathtools}
\usepackage{footmisc}

\usepackage{tikz}
\usepackage{pgfplots}
\pgfplotsset{compat=1.16}

\usepgfplotslibrary{colorbrewer}

\usepackage{newtxtext,newtxmath}

\numberwithin{equation}{section}

\newcommand*{\scalarproduct}[2]{\left(\,#1\,\middle|\,#2\,\right)}
\newcommand*{\operator}[1]{\hat{#1}}
\newcommand*{\complexi}{\mathrm{i}}
\newcommand*{\transpose}{\mathbfss{T}}

\title[Operators and the spherical harmonic expansion]{Using Spherical Harmonics to solve the Boltzmann equation: an operator based approach}

\author[Nils W. Schween et al.]{
Nils W. Schween,$^{1}$\thanks{E-mail: nils.schween@mpi-hd.mpg.de}
Brian Reville,$^{1}$
\\
$^{1}$Max-Planck-Institut f\"{u}r Kernphysik, Astrophysical Plasma Theory (APT),
Saupfercheckweg 1, 69117 Heidelberg, Germany\\
}

\date{Accepted 2024 February 23. Received 2024 February 20; in original form 2024 January 30}

\pubyear{2024}

\begin{document}
\label{firstpage}
\pagerange{\pageref{firstpage}--\pageref{lastpage}}
\maketitle

\begin{abstract}
  The transport of charged particles or photons in a scattering medium can be modelled with a
  Boltzmann equation. The mathematical treatment for scattering in such scenarios is often
  simplified if evaluated in a frame where the scattering centres are, on average, at rest. It is
  common therefore, to use a mixed coordinate system, wherein space and time are measured in a
  fixed inertial frame, while momenta are measured in a ``co-moving'' frame. To facilitate
  analytic and numerical solutions, the momentum dependency of the phase-space density may be
  expanded as a series of spherical harmonics, typically truncated at low order. 
  A method for deriving the system of
  equations for the expansion coefficients of the spherical harmonics to arbitrary order is presented in
  the limit of isotropic, small-angle scattering. The method of derivation takes advantage of
  operators acting on the space of spherical harmonics. The matrix representations of these
  operators are employed to compute the system of equations. The computation of matrix
  representations is detailed and subsequently simplified with the aid of rotations of
  the coordinate system. The eigenvalues and eigenvectors of the matrix representations are
  investigated to prepare the application of standard numerical techniques, e.g. the finite volume
  method or the discontinuous Galerkin method, to solve the system.
\end{abstract}




\section{Introduction}
\label{sec:introduction}

The transport of charged particles, photons or neutrinos in an inhomogeneous scattering medium is a 
recurring problem that appears across multiple sub-fields of physics \cite[e.g.][]{Braginskii,Chandrasekhar, Ginzburg, Mihalas, lewis84}. Astrophysical contexts include neutrino and radiation hydrodynamics in supernova explosions, as well as
the propagation and acceleration of energetic charged particles (cosmic rays).
Typically, one considers the evolution of the single particle phase-space density,
$f(\mathbfit{x}, \mathbfit{p}, t)$ by solving a transport
equation which can be written in the form \( \frac{\textrm{d} f}{\textrm{d} s} = \hat{C}f\, \). Here \(s\) is an affine parameter that defines a point on a photon
or particle trajectory (for a photon or neutrino the distance along a ray, for a particle with mass the proper
time) and \(\hat{C}\) is a collision operator that describes stochastic disturbances of these
trajectories. For photons and neutrinos, the trajectories are simply geodesics in the local space-time, and
\(\hat{C}\) contains the scattering, absorption and emission cross-sections. These may be prescribed in the scattering-centres' rest frame, which may be determined by solving a separate set of equations (e.g., those of hydro- or magnetohydrodynamics). These equations in turn can be augmented by terms containing
moments of \(f\), i.e. feedback. In general, the rest frame of the scattering centres is non-inertial, and this must be taken into account when formulating the relevant transport equation \cite[e.g.][]{Achterberg2018_1}. 

For example, the phase-function for photon scattering is usually specified in terms of the
coordinates of the momenta \(\mathbfit{p}\) and \(\mathbfit{p}'\) of the photons involved, measured
in the ``co-moving'' frame in which the scattering centres are locally at rest. This leads
to a formulation of the radiation transport equation in so-called ``mixed'' coordinates
\citep{Lindquist66, Castor72, Riffert86}, with the phase-space density treated as a function of
position \(\mathbfit{x}\) and time \(t\), both measured in the laboratory frame, while the momentum
\(\mathbfit{p}\), is measured in the co-moving frame.

For cosmic rays, the situation is more complicated. As in the case of photons and neutrinos, one conventionally assumes the presence of a ``background''
plasma whose evolution can be determined, for example via the equations of ideal magnetohydrodynamics. But, in this case, these are augmented not only by terms in the stress-energy tensor, but also by the current and charge density carried by the cosmic rays. The charged particle trajectories are then not geodesics, but are determined by the electromagnetic fields carried by the background plasma.
In dense plasmas,
where Coulomb collisions among the different charged species occur, \(\hat{C}\) can be modelled in the
co-moving frame as a Fokker-Planck operator \citep{RMJ57,Shkarofsky_ParticleKineticsOfPlasmas}. However, in dilute, astrophysical plasmas, Coulomb collisions are unimportant,
and \(\hat{C}\) is instead used to model the effects of scattering on stochastic turbulent fluctuations that are 
superimposed on the large-scale fields carried by the background plasma. Given suitable constraints on the properties of this turbulence, \(\hat{C}\) can again be formulated as a Fokker-Planck operator, this time in the (non-inertial) \emph{wave-frame} \cite[e.g.][]{Skilling1975, Webb85,
  Kirk88}\footnote{There is an erratum to \citet{Webb85}, which is given by the
author in \citet{Webb1987}, where it is noted that the advective transport term in
equation (8.5) should be multiplied by the Lorentz factor of the flow.}.
In principle, the large scale fields take into account the current and charge density of the cosmic rays. Consequently, the left-hand-side of the transport equation resembles a Vlasov equation, which has given rise to the name
``Vlasov--Fokker--Planck equation''.

For both photons and charged particles, the single particle phase-space density is a function of six
independent variables and time, making direct discretisation approaches prohibitive.
In the case of the space and time variables, numerical techniques for solving the MHD equations are well-known. These can
be applied also to the VFP equation, since \(f\) is not expected to vary on scales shorter or faster than those of the MHD variables.
Also, CR distributions are generally smooth functions of the particle energy (or the modulus \(p\) of the momentum), so that discretization
of this variable is not problematic. However, the two remaining angular variables (in momentum space) pose a problem, since strong anisotropies can develop, in
particular close to oblique shock fronts in the background plasma or in the presence of highly relativistic flows. 
An effective approach here is
to expand 
the phase-space density in a series of known functions, and thereby resolve the anisotropy by extending the expansion to higher order until convergence is achieved
\citep{Kirk87, Kirk2023}. In the case of cosmic rays in non-relativistic flows, a suitable set of functions is that of spherical
harmonics. This is the approach we adopt here, by expanding the single particle distribution in a series of spherical harmonics \(Y^{m}_{l}\), which are functions
of the spherical polar coordinates \(\theta\) and \(\phi\) of the momentum vector \(\mathbfit{p}\) (defined with axis along \(p_x\)):
\begin{equation}
	\label{eq:spherical-harmonic-exp}
	f(\mathbfit{x},p, \theta, \varphi, t) =
	\sum^{\infty}_{l = 0} \sum^{l}_{m = -l } f^{m}_{l}(\mathbfit{x}, p, t) Y^{m}_{l}(\theta, \varphi) \,,
\end{equation}
or, equivalently, using Cartesian tensors \citep{Schween2022}. The equivalence is particularly well exploited in the
(projected-)symmetric trace-free formalism described by \citet{Thorne1980,
  Thorne1981}. Alternatively, it is possible to derive the equations for the moments of the
distribution function and, thus, to directly investigate the relevant transport
coefficients, see for example \citet{Webb1989} and references therein.
Expanding the single particle distribution function reduces the number of dependent variables from six to four, though it requires a procedure to determine an effective transport equation for each coefficient in the expansion. While derivations of this system of equations can be found in the literature for different applications, both for Cartesian expansions \cite[see][truncated at low order]{Johnston1960, WilliamsJokipii91, Thomas2012} and spherical harmonics \cite[e.g.][]{Bell2006,Tzoufras2011,Reville2013}, the resulting equations are cumbersome, and the physical meaning is obfuscated. 

In this paper we present a simplification of this expansion procedure which avoids lengthy algebraic
manipulations, when expanding to high order. This facilitates
checking the convergence of the expansion and allows for a direct physical interpretation at each stage. We restrict our derivation
to the propagation and acceleration of charged particles in a prescribed background plasma. In
Section~\ref{sec:particle-acceleration} we introduce the relevant transport equation, namely a
variant of the Vlasov--Fokker--Planck (VFP) equation, and simplify it by making the additional,
although not necessary, assumption that the co-moving frame moves slowly in the laboratory frame,
and that the scattering operator is isotropic. For such a scattering operator, the eigenfunctions
are the spherical harmonics used in the expansion. This motivates an approach which exploits
operators acting on the space of spherical harmonics and their matrix representations. In
Section~\ref{sec:derivation-of-the-system-of-equations} we identify and define the operators which
are present in the VFP equation and show that the system of equations can then be obtained by
replacing the operators with their matrix representations. This has the advantage of retaining the
original structure of the VFP equation. We explicitly compute the matrix elements of the matrix
representations. Moreover, in Section~\ref{sec:rotations-spherical-harmonics} we prove that the
effort of computing them can be considerably reduced with the help of rotations of the coordinate
system in momentum space. We investigate how these rotations act in the space of spherical harmonics
and how they change the operators and their matrix representations. The key insight is that the
operators' matrix representations can be considered as ``rotated'' versions of each other and thus it
is possible to compute any one of them by rotating another. We quote explicit expressions for the
necessary rotation matrices. Having derived and simplified the computation of the system of
equations, we note that it contains redundant equations. The reason is that the reality of the single particle
phase-space density imposes symmetries on the expansion coefficients. In
Section~\ref{sec:real-system-of-equations} we remedy this using the \emph{real} spherical harmonics.
The relation between the complex and real spherical harmonics can be expressed with the help of a
basis transformation matrix. Multiplying the system of equations with this basis transformation
matrix removes the superfluous equations. Numerically solving the system of equations using standard
techniques, e.g. the finite volume method or the discontinuous Galerkin method, requires knowledge
about the eigenvalues and eigenvectors of the matrix representations. In the last part of this
paper, i.e . in Section~\ref{sec:eigenvectors-and-eigenvalues-representation matrices}, we prove
that the representation matrices have the same eigenvalues and that there eigenvectors are also
rotated versions of each other. Moreover, we show that eigenvalues are the roots of the associated
Legendre polynomials. The implementation of the discontinuous Galerkin method exploiting this
knowledge is the subject of a companion paper.

\section{Particle transport in plasmas}
\label{sec:particle-acceleration}

The transport of charged particles and photons in a scattering, non-stationary medium was recently revisited by \cite{Achterberg2018_1}. Their work provides a rigorous derivation of the mixed-frame transport equation in the context of special relativity. We adopt the mixed-frame transport equation as our starting point, and refer the reader to their paper for further details.  

We consider particles that move in a background fluid, whose velocity field can be described with the 4-vector \(U^\mu =\Gamma(c,\mathbfit{U})\) (bold symbols denoting 3-vectors), with \(\Gamma = (1-U^2/c^2)^{-1/2}\) the bulk Lorentz factor of the background fluid. For the particles on the other hand, the four-velocity is \(V^\mu =\gamma(c,\mathbfit{V})\), with $\gamma = p/mV= \varepsilon/m c^2$ the associated particle Lorentz factor. The background plasma is permeated by electric and magnetic fields, $\mathbfit{E}$ and $\mathbfit{B}$. Using unprimed quantities to denote quantities measured in the laboratory frame, and primed quantities to denote those measured in the fluid rest frame, one arrives at the following transport equation
\begin{equation}
\label{eq:vfp-mixed-coordinates-rel}
\frac{d f}{d s} + \left[\gamma'  q\left(\mathbfit{E}' + \mathbfit{V}'  \times \mathbfit{B}' \right) - 
(\Gamma+1) \varepsilon' \frac{d}{ds}\left(\frac{\Gamma \mathbfit{U}}{\Gamma+1}\right) + \mathbfit{p}'\times \bomega_\ell \right]\cdot \nabla_{p'} f
= \gamma' \frac{\nu'}{2}\Delta_{\theta', \varphi'}f \,,
\end{equation}
where, on account of the mixed coordinates, the proper-time derivative along the particle trajectory is
\begin{equation}
\label{eq:proper-time-derivative}
\frac{d \,}{d s} = \frac{\,p^\nu}{m}\frac{\partial \,}{\partial x^\nu} = 
\gamma'\left\lbrace \Gamma\left(1 + \frac{\mathbfit{V}' \cdot \mathbfit{U}}{c^{2}} \right)\frac{\partial \,}{\partial t} + \left(\Gamma \mathbfit{U} +
\mathbfit{V}'\right) \cdot \nabla_{x} +\frac{\Gamma^2}{\Gamma+1}\frac{\mathbfit{V}'\cdot\mathbfit{U}}{c^2}(\mathbfit{U}\cdot\nabla_{x}) \right\rbrace \,.
\end{equation}
As described in \citet[eq. 15]{Achterberg2018_1}, the second two terms inside the square brackets of equation (\ref{eq:vfp-mixed-coordinates-rel}) correspond to the fictitious forces introduced from the accelerating frame. The $\bomega_\ell$ vector stems from changes in the  \emph{direction} of $\mathbfit{U}$, 
\begin{equation*}
 \bomega_\ell = (\Gamma - 1) \left(\bm{\hat{\ell}} \times \frac{\mathrm{d} \bm{\hat{\ell}}}{\mathrm{d} s} \right) \quad\text{with } \bm{\hat{\ell}} \text{ a unit vector along } \mathbfit{U} = U \bm{\hat{\ell}}\, .
\end{equation*}
The particular form of the collision term we adopt here corresponds to isotropic scattering in the local fluid frame with rate $\nu'(p')$,  although such a simplification is not essential. The symbol
\(\Delta_{\theta', \varphi'}\) denotes the Laplace operator in spherical angles,
\begin{equation}
 \label{eq:laplace-angular-part}
 \Delta_{\theta', \varphi'}
= \left(\frac{\partial^{2}}{\partial \theta'^{2}}
   + \frac{\cos\theta'}{\sin\theta'} \frac{\partial}{\partial \theta'}
   + \frac{1}{\sin^{2}\theta'} \frac{\partial^{2}}{\partial \varphi'^{2}}
 \right) \,.
\end{equation}
The collision term thus assumes that the scattering is elastic, changing only the direction of a particle's momentum and not its magnitude which need not be the case if the scattering centres are dispersive, or anisotropic \citep{Melrose69,Skilling1975}. A detailed study of these effects is beyond the scope of the current paper.

For $U \ll c$, we can expand equation \eqref{eq:vfp-mixed-coordinates-rel} retaining all terms to first order in $U/c$:
\begin{equation}
\label{eq:vfp-mixed-coordinates}
\left(1 + \frac{\mathbfit{V}' \cdot \mathbfit{U}}{c^{2}} \right)\frac{\partial f}{\partial t} + \left(\mathbfit{U} +
\mathbfit{V}'\right) \cdot \nabla_{x} f
+ \left[\left( q\mathbfit{E}'  -\gamma' m \frac{\mathrm{d} \mathbfit{U}}{\mathrm{d} t}\right)
- (\mathbfit{p}' \cdot \nabla_{x})\mathbfit{U} + q\mathbfit{V}'  \times \mathbfit{B}'  \right]\cdot \nabla_{p'} f
= \frac{\nu'}{2}\Delta_{\theta', \varphi'}f \,.
\end{equation}

In the rest of this text we will work with eq.~\eqref{eq:vfp-mixed-coordinates}, but we emphasise that
the method to derive the equations for the expansion coefficients of the spherical harmonics can be
straightforwardly applied to the fully relativistic VFP equation (\ref{eq:vfp-mixed-coordinates-rel}). 
We will henceforth drop
the primes from all co-moving terms, and it is to be understood that 
$\mathbfit{U},\mathbfit{V}, \mathbfit{p},\gamma, \mathbfit{B}$ and $\nu$ are
referring to quantities defined in the rest frame of the thermal plasma. 
We also restrict our attention to ideal MHD, 
for which $\mathbfit{E}'=0$, though since it independent of $\mathbfit{V}$, it can be re-introduced 
if required by adding to the $d\mathbfit{U}/dt$ term. 

\section{Derivation of the system of equations}
\label{sec:derivation-of-the-system-of-equations}

We now convert eq.~\eqref{eq:vfp-mixed-coordinates} into a matrix equation 
by inserting the expansion \eqref{eq:spherical-harmonic-exp} and projecting onto the spherical harmonics. 
Projection involves integrals over the unit sphere. We adopt a shorthand notation for
the corresponding scalar product:
\begin{equation}
  \label{eq:scalar-product-sphere}
  \scalarproduct{f}{g} \coloneqq \int_{S^{2}} {f}^{*}(\theta, \varphi) g(\theta, \varphi)  \mathrm{d}\Omega \, .
\end{equation}
This yields
\begin{equation}
  \label{eq:vfp-mixed-spherical-exp}
  \begin{split}
    &\sum^{\infty}_{l = 0} \sum^{l}_{m = -l}
    \scalarproduct{Y^{m'}_{l'}}{Y^{m}_{l}} \, \frac{\partial f^{m}_{l}}{\partial t}
    + \frac{1}{c^{2}}
    \scalarproduct{Y^{m'}_{l'}}{\mathbfit{V} \cdot \mathbfit{U} Y^{m}_{l}} \frac{\partial f^{m}_{l}}{\partial t}  
    +
    \scalarproduct{Y^{m'}_{l'}}{\left(\mathbfit{U} + \mathbfit{V}\right)Y^{m}_{l}}\cdot \nabla_{x}f^{m}_{l}\\
    &\qquad
      {} -
      \scalarproduct{Y^{m'}_{l'}}{\left(\gamma m \frac{\mathrm{d} \mathbfit{U}}{\mathrm{d} t}
      + (\mathbfit{p} \cdot \nabla_{x}) \mathbfit{U} \right) \cdot \nabla_{p} (f^{m}_{l}Y^{m}_{l})}
      +
      q\scalarproduct{Y^{m'}_{l'}}{\mathbfit{V} \cdot \left(\mathbfit{B} \times \nabla_{p}( f^{m}_{l}Y^{m}_{l} )\right)}\\
      &\qquad =  \frac{\nu}{2} \sum^{\infty}_{l = 0} \sum^{l}_{m = -l}
      \scalarproduct{Y^{m'}_{l'}}{\Delta_{\theta, \varphi}Y^{m}_{l}}\,f^{m}_{l}
      \quad \text{for all } l' \in \mathbb{N}_{0} \text{ and } |m'| \leq l' \,. 
  \end{split}
\end{equation}
Eq.~\eqref{eq:vfp-mixed-spherical-exp} can be interpreted as a matrix equation. The scalar products
are the elements of matrices that multiply the vector
$\left(\, \mathbfit{f}\right)_{j(l,m)} \coloneqq f^{m}_{l}$ or its derivatives. The index $j$ used
here is a one-to-one function of the indices $l$ and $m$. We call this function an \textit{index
  map} and it determines how the expansion coefficients are ordered. An example order is given
through $\mathbfit{f} = (f^{0}_{0}, f^{1}_{1}, f^{0}_{1}, f^{-1}_{1}, \dots)^{T}$. In this case the
index map $j$ is $j(l, m) = l(l + 1) - m$, with $j$ starting at zero. Its (unique) inverse is
given as $l = \lfloor \sqrt{j}\rfloor$ and $m = l(l+1)-j$.

\subsection{The identity operator and the collision operator}
\label{sec:identitiy-collision-operator}

As a first example, we consider the time derivative. This can be written in the form:
\begin{equation}
  \label{eq:time-derivative-term}
  \sum^{\infty}_{l = 0} \sum^{l}_{m = -l} \scalarproduct{Y^{m'}_{l'}}{Y^{m}_{l}} \frac{\partial f^{m}_{l}}{\partial t}
  = \sum^{\infty}_{l = 0} \sum^{l}_{m = -l} (\mathbfss{1})_{i(l',m') j(l,m)}\partial_{t}(\, \mathbfit{f})_{j(l,m)} \, ,
\end{equation}
where $\mathbfss{1}$ is the identity matrix. Note that we introduced a second index map $i(l', m') = l'(l' + 1) - m'$ such that the notation resembles ordinary  matrix--vector product. In particular, we emphasise that summing over $l$ and $m$ can be interpreted as the multiplication of the identity matrix with the time derivative of the vector $\mathbfit{f}$. Each time additional matrix indices are required a ``copy'' of the index map $i$ (or $j$) will be used, see for example the matrix-matrix products in eq. \eqref{eq:matrix-represenation-of-product-of-operators}.

The identity matrix can be considered to be the \textit{matrix representation} of the identity
operator $\operator{1}$ acting on the space of the spherical harmonics. We define the matrix
representation of an operator $\operator{O}$ acting on an inner product space
$\mathcal{H} = \text{span}\{b_{1}, b_{2}, \dots\}$ to be 
\begin{equation}
  \label{eq:def-matrix-representation}
  (\mathbfss{O})_{ij} = \scalarproduct{b_{i}}{\operator{O}b_{j}}
  \quad\text{ where the } b_{i}\text{s are the basis vectors of } \mathcal{H}\,.
\end{equation}
We note that the space of all spherical harmonics is a Hilbert space, whose inner product is defined by the orthogonality relation
 \begin{equation}
  \label{eq:orthogonality-spherical-harmonics}
  \int_{S^{2}} {Y^{m'}_{l'}}^{*} Y^{m}_{l} \mathrm{d}\Omega = \delta_{l'l} \delta_{m'm} \quad
  \text{where $S^{2}$ is the unit sphere, i.e. }
  S^{2} \coloneqq \{\mathbfit{x} \in \mathbb{R}^{3} \mid x \coloneqq \|x\| = 1\} \,.
\end{equation}
We denote this space by
\begin{equation}
  \label{eq:space-spherical-harmonics}
  \mathcal{S} \coloneqq \text{span}\{Y^{0}_{0}, Y^{1}_{1}, Y^{0}_{1}, Y^{-1}_{1}, \dots\} \,.
\end{equation}

The collision operator serves as another example for the interpretation of the integrals as
matrix elements. The right-hand side of eq.~\eqref{eq:vfp-mixed-spherical-exp} is
\begin{equation}
  \label{eq:collision-term}
  \frac{\nu}{2} \sum^{\infty}_{l = 0} \sum^{l}_{m = -l} \scalarproduct{Y^{m'}_{l'}}{\Delta_{\theta, \varphi}Y^{m}_{l}} f^{m}_{l}
  \coloneqq -\nu \sum^{\infty}_{l = 0} \sum^{l}_{m = -l} (\mathbfss{C})_{i(l',m') j(l,m)} (\, \mathbfit{f})_{j(l,m)} \,.
\end{equation}
Where $\mathbfss{C}$ is the matrix representation of the collision operator in the
spherical harmonic basis. It is a diagonal matrix and its elements are easily computed, since the
spherical harmonics are eigenfunctions of the angular part of the Laplace operator, see
eq.~\eqref{eq:laplace-angular-part}. They are
\begin{equation}
  \label{eq:matrix-representation-collision}
  (\mathbfss{C})_{i(l',m') j(l,m)}
  = -\frac{1}{2} \scalarproduct{Y^{m'}_{l'}}{\Delta_{\theta, \varphi}Y^{m}_{l}}
  = \frac{l(l + 1)}{2} \delta_{ll'}\delta_{m'm} \,.
\end{equation}

These two examples suggest that the
system of equations which determine the expansion coefficients $f^{m}_{l}$ can be formulated as
representation matrices of operators acting on the space of spherical harmonics. We thus focus on
identifying other operators whose action on the spherical harmonics are either known or easily 
derived.

\subsection{The angular momentum operator}
\label{sec:angular-momentum-operator}

The magnetic force term in the VFP equation~\eqref{eq:vfp-mixed-coordinates} is the counterpart in momentum space
of the \textit{angular momentum operator} in configuration space that is well-known from quantum mechanics:
$\mathbfit{L}_{\text{QM}} \coloneqq -\complexi \mathbfit{x} \times \nabla_{x}$. I.~e.,
\begin{equation}
  \label{eq:magnetic-force}
  q \mathbfit{V} \cdot \left(\mathbfit{B} \times \nabla_{p}f\right)
  = - q  \mathbfit{B} \cdot \left(\mathbfit{V}  \times  \nabla_{p}f \right)
  = -\frac{q \mathbfit{B}}{\gamma m} \cdot \left(\mathbfit{p} \times \nabla_{p}\right) f
  \coloneqq - \complexi \bomega \cdot \mathbfit{L}f  \,,
\end{equation}
where $\bomega= q \mathbfit{B}/\gamma m$ is the angular frequency vector, and $\mathbfit{L}=-\complexi \mathbfit{p} \times \nabla_{p}$.
We choose to keep the name ``angular momentum operator'' for $\mathbfit{L}$, despite its action on the momentum space variables $\mathbfit{p}$

In terms of the coordinates \(\theta\) and \(\phi\),
\begin{equation}
  \begin{split}
    \label{eq:angular-momentum-operators}
    \operator{L}_{x}     & \coloneqq -\complexi \partial_{\varphi}\quad \text{and }  \\
    \operator{L}_{\pm} & \coloneqq \operator{L}_{y} \pm \complexi \operator{L}_{z}
                         = e^{\pm \complexi \varphi}\left(\pm \frac{\partial}{\partial \theta} + \complexi \cot\theta \frac{\partial}{\partial \varphi} \right) \,.  
  \end{split}
\end{equation}
The operator $\operator{L}_{\pm}$ is the \textit{angular momentum ladder operator}. These
definitions can for example be found in \citet[eq. 26.14--15]{Landau_QuantumMechanics}. We express
$\operator{L}_{y}$ and $\operator{L}_{z}$ in terms of these ladder operators and we get for the
angular momentum operator
\begin{equation}
  \label{eq:L-operator-ladder-operators}
  \mathbfit{L} =
  \begin{pmatrix}
    \operator{L}_{x} \\
    1/2 \left(\operator{L}_{+} + \operator{L}_{-}\right) \\
    1/2\complexi \left(\operator{L}_{+} - \operator{L}_{-}\right)
  \end{pmatrix} \,.
\end{equation}

We stress that the angular momentum operator acts only on $\theta$ and $\phi$. Hence, the magnetic force term in eq.~\eqref{eq:vfp-mixed-spherical-exp} becomes
\begin{equation}
  \label{eq:angular-momentum-op-representation-matrices}
  \int_{S^{2}} {Y^{m'}_{l'}}^{*} q \mathbfit{V} \cdot \left(\mathbfit{B} \times \nabla_{p}( f^{m}_{l}Y^{m}_{l} )\right)\, \mathrm{d}\Omega
  = - \complexi \bomega \cdot \scalarproduct{Y^{m'}_{l'}}{\mathbfit{L} Y^{m}_{l}} f^{m}_{l}
  \coloneqq -\complexi \omega_{a}(\bm{\Omega}^{a})_{i(l',m')j(l, m)}(\, \mathbfit{f})_{j(l, m)} \,,
\end{equation}
where we took $f^{m}_{l}$ out of the integral and introduced the representation matrices $\bm{\Omega}^{a}$ of the angular momentum operators.

Note that we implicitly sum over the index $a \in \{x, y, z\}$. We will frequently write $\bm{\Omega}^{a}$ to refer to all three representation matrices at once. The action of the angular momentum operators on the
spherical harmonics is well-known from quantum mechanics. For example, in \citet[\S 27]{Landau_QuantumMechanics}
their action is given to be
\begin{equation}\label{eq:action-L-operators}
  \begin{split}
    \operator{L}_{x}Y^{m}_{l} &= m Y^{m}_{l} \, ,  \\
    \operator{L}_{+}Y^{m}_{l} & = \sqrt{(l + m + 1)(l - m )} Y^{m + 1}_{l}\quad \text{and} \\
    \operator{L}_{-}Y^{m}_{l} & = \sqrt{(l + m )(l - m + 1)} Y^{m - 1}_{l} \,.
  \end{split}
\end{equation}

The matrix representations of the angular momentum operators can now be computed using
eq.~\eqref{eq:action-L-operators} and eq. \eqref{eq:L-operator-ladder-operators}. They are
\begin{align}
  \label{eq:Omegax}
  (\bm{\Omega}^{x})_{i(l',m')j(l,m)} &= m \delta_{l'l} \delta_{m'm} \, , \\
  \label{eq:Omegay}
  (\bm{\Omega}^{y})_{i(l',m')j(l,m)} &=
                                       \frac{1}{2} \sqrt{(l + m + 1)(l - m)} \delta_{l'l}\delta_{m'(m+1)} + \frac{1}{2} \sqrt{(l + m) (l - m + 1)} \delta_{l'l} \delta_{m'(m-1)} \, ,\\
  \label{eq:Omegaz} 
  (\bm{\Omega}^{z})_{i(l', m')j(l,m)} &= -\frac{\complexi}{2} \sqrt{(l + m + 1)(l - m)} \delta_{l'l}\delta_{m'(m+1)} + \frac{\complexi}{2} \sqrt{(l + m) (l - m + 1)} \delta_{l'l} \delta_{m'(m-1)} \,.
\end{align}
Alternatively, these matrix elements can also be found in \citet[Chapter 13.2, eq.
42]{Varshalovich_QuantumTheoryOfAngularMomentum}.

\subsection{The direction operators}
\label{sec:direction-operators}

In addition to the identity and angular momentum operators, another set of operators appears in the
VFP equation. We call them
\textit{direction operators} because they ``point'' in the direction of the coordinate axes:
\begin{equation}
  \label{eq:A-operators}
  \mathbfit{V} = V
  \begin{pmatrix}
    \cos\theta \\
    \sin\theta \cos\varphi \\
    \sin\theta \sin\varphi \\
  \end{pmatrix}
  \coloneqq V
  \begin{pmatrix}
    \operator{A}^{x} \\
    \operator{A}^{y} \\
    \operator{A}^{z}
  \end{pmatrix} \,.
\end{equation}
These operators act on the space of spherical harmonics, i.e.
$\operator{A}^{a}: \mathcal{S} \rightarrow \mathcal{S}$. 
They appear, for example, in the part of the VFP equation we call the
\textit{spatial advection term}. In the equations for the expansion
coefficients~\eqref{eq:vfp-mixed-spherical-exp}, the spatial advection term is
\begin{equation}
  \label{eq:advection-term}
  \begin{split}
    \int_{S^{2}} {Y^{m'}_{l'}}^{*}\left(\mathbfit{U} + \mathbfit{V}\right)Y^{m}_{l} \, \mathrm{d}\Omega \cdot \nabla_{x}f^{m}_{l}  &= \left(U^{a}\scalarproduct{Y^{m'}_{l'}}{Y^{m}_{l}}
    + V \scalarproduct{Y^{m'}_{l'}}{\operator{A}^{a}Y^{m}_{l}}\right)\frac{\partial f^{m}_{l}}{\partial x^{a}} \\
    &\coloneqq \left(U^{a}(\mathbfss{1})_{i(l', m')j(l,m)} + V (\mathbfss{A}^{a})_{i(l', m')j(l,m)} \right) \partial_{x_{a}}(\, \mathbfit{f})_{j(l,m)} \,.
  \end{split}
\end{equation}
They also appear in front of the time derivative of the distribution function, i.e.
\begin{equation}
  \label{eq:time-correction-term}
  \frac{1}{c^{2}} \int_{S^{2}} {Y^{m'}_{l'}}^{*} \mathbfit{V} \cdot \mathbfit{U} Y^{m}_{l} \, \mathrm{d}\Omega \frac{\partial f^{m}_{l}}{\partial t}
  = \frac{V}{c^{2}} U_{a}\scalarproduct{Y^{m'}_{l'}}{\operator{A}^{a}  Y^{m}_{l}} \frac{\partial f^{m}_{l}}{\partial t}
  = \frac{V}{c^{2}} U_{a} (\mathbfss{A}^{a})_{i(l',m')j(l,m)}\partial_{t}(\mathbfit{f} \,)_{j(l,m)} \,.
\end{equation}

As before, the computation of the matrix representations of the $\operator{A}^{a}$ operators leads us
to investigate their action on the spherical harmonics. There are multiple ways to determine their
action. For example, it is possible to use recurrence relations for the associated Legendre
polynomials or to use ladder operators, which increase/decrease $l$ and/or $m$, to represent them.

We begin with $\operator{A}^{x}$ and we use a recurrence relation to examine its action, namely
\begin{equation}
  \label{eq:recurrence_relation_xP}
  \cos\theta P^{m}_{l}(\cos\theta) =
  \frac{l - m + 1}{2l + 1} P^{m}_{l+1}(\cos\theta) + \frac{l + m}{2l + 1} P^{m}_{l-1}(\cos\theta) \,.
\end{equation}
With its help, we show that the action of $\operator{A}^{x}$ is
\begin{equation}
  \label{eq:recurrence_relation_xY}
  \begin{split}
    \operator{A}^{x} Y^{m}_{l} = \cos\theta Y^{m}_{l} &=
                           \frac{l - m + 1}{2l + 1} \frac{N^{m}_{l}}{N^{m}_{l + 1}} N^{m}_{l+1} P^{m}_{l+1} e^{\complexi m \varphi}
                           + \frac{l + m}{2l + 1} \frac{N^{m}_{l}}{N^{m}_{l - 1}} N^{m}_{l-1} P^{m}_{l-1}(\cos\theta)e^{\complexi m \varphi}  \\
                         &= \sqrt{\frac{(l + m + 1)(l - m + 1)}{(2l + 3)(2l + 1)}} Y^{m}_{l+1}
                         + \sqrt{\frac{(l + m)(l - m)}{(2l + 1)(2l - 1)}} Y^{m}_{l-1} \,.
  \end{split}
\end{equation}

Instead of using a recurrence relation for the associated Legendre polynomials, we can work with
ladder operators which increase (decrease) $l$. These operators are investigated in \citet[eq.
12]{Fakhri2016} and they are defined as
\begin{equation}
  \label{eq:l-ladder-operator}
  \operator{J}_{\pm}(l) \coloneqq \pm \sin\theta + l \cos\theta \,.
\end{equation}
With this definition at hand, a short computation shows that
\begin{equation}
  \label{eq:Ax-shift-operators}
  \operator{A}^{x} = \frac{1}{2l + 1} \left(\operator{J}_{+}(l + 1) + \operator{J}_{-}(l)\right) \,.
\end{equation}

The action of $\operator{A}^{x}$ is now given by the action of the ladder operators
$\operator{J}_{\pm}$, which can as well be found in \citet[eq. 14 a,b]{Fakhri2016}, and it is
\begin{align}
  \label{eq:action-J}
  \operator{J}_{+}(l + 1) Y^{m}_{l} &= \sqrt{\frac{2l + 1}{2l +3} (l - m + 1) (l + m + 1)} Y^{m}_{l + 1}\quad \text{and} \\
  \operator{J}_{-}(l) Y^{m}_{l} &= \sqrt{\frac{2l + 1}{2l  - 1} (l - m) (l + m)} Y^{m}_{l + 1}  \,.
\end{align}

For the sake of completeness, we proceed analogously with the operator $\operator{A}^{y}$ and
$\operator{A}^{z}$, i.e. we, first, compute their action using a suitable recurrence relation for
the associated Legendre polynomials and, secondly, we express their action with the help of
appropriate ladder operators. But already at this point, we highlight that an
explicit computation of the representation matrices of $\operator{A}^{y}$ and $\operator{A}^{z}$ is
not necessary, as they are connected via rotations (see section \ref{sec:change-operator-rotation}).

To compute the action of $\operator{A}^{y}$ and $\operator{A}^{z}$, we use the two recurrence relations
\begin{equation}
  \label{eq:recurrence-relation-sinP}
  \begin{split}
    \sin\theta P^{m}_{l}(\cos\theta) &= \frac{1}{2l + 1}
                                       \left[(l - m + 1)(l - m + 2) P^{m - 1}_{l + 1} - (l + m - 1)(l + m)P^{m-1}_{l+1}\right] \quad \text{and} \\
    \sin\theta P^{m}_{l}(\cos\theta) &= \frac{-1}{2l + 1} \left[P^{m + 1}_{l + 1} - P^{m+1}_{l-1}\right] \,.
  \end{split}
\end{equation}
With these relations at hand, we show that the action of $\operator{A}^{y}$ is
\begin{equation}
  \label{eq:action-Ay}
  \begin{split}
    \operator{A}^{y}Y^{m}_{l} &= \sin\theta \cos\varphi Y^{m}_{l} = \frac{1}{2} \sin\theta \left(e^{\complexi \varphi} + e^{-\complexi \varphi}\right) Y ^{m}_{l}
                           = \frac{1}{2} N^{m}_{l}\left(\sin\theta P^{m}_{l} e^{\complexi (m + 1) \varphi}
                           + \sin\theta P^{m}_{l} e^{\complexi (m - 1)\varphi} \right) \\
                         &= \frac{1}{2(2l + 1)}\left(- \frac{N^{m}_{l}}{N^{m+1}_{l + 1}} Y^{m+1}_{l+1} + \frac{N^{m}_{l}}{N^{m+1}_{l - 1}} Y^{m+1}_{l - 1} + (l - m + 1)(l - m + 2) \frac{N^{m}_{l}}{N^{m-1}_{l +1}} Y^{m-1}_{l+1} - (l + m + 1)(l + m) \frac{N^{m}_{l}}{N^{m-1}_{l-1}} Y^{m-1}_{l-1}\right) \\
    &= \frac{1}{2}\left(- a^{m+1}_{l+1}Y^{m+1}_{l+1} +  a^{-m}_{l} Y^{m+1}_{l - 1} +  a^{-m + 1}_{l+1} Y^{m-1}_{l+1} - a^{m}_{l} Y^{m-1}_{l-1}\right) \, .
  \end{split}
\end{equation}
The coefficient $a^{m}_{l}$ is 
\begin{equation}
  \label{eq:coefficients-action-Ay-Az}
    a^{m}_{l} = \sqrt{\frac{(l + m)(l + m - 1)}{(2l + 1)(2l - 1)}} \, .
\end{equation}
In a completely analogous manner the action of $\operator{A}^{z}$ can be shown to be
\begin{equation}
  \label{eq:action-Az}
  \begin{split}
    \operator{A}^{z}Y^{m}_{l} &= \sin\theta \sin\varphi Y^{m}_{l}
                                = \frac{1}{2\complexi} \sin\theta \left(e^{\complexi \varphi} - e^{-\complexi \varphi}\right) Y ^{m}_{l}
                                = \frac{\complexi}{2}\left( a^{m+1}_{l+1}Y^{m+1}_{l+1} -  a^{-m}_{l} Y^{m+1}_{l - 1} +  a^{-m+1}_{l+1} Y^{m-1}_{l+1} - a^{m}_{l} Y^{m-1}_{l-1}\right) \, .
  \end{split}
\end{equation}
Again, the action of $\operator{A}^{y}$ and $\operator{A}^{z}$ can alternatively be derived with
ladder operators which can be found in \citet[eq. 25, eq. 34]{Fakhri2016}, i.e.
\begin{align}
  \label{eq:lm_shift_operators}
  \operator{A}^{\pm}_{\pm}(l) &\coloneqq \pm [\operator{L}_{\pm}, \operator{J}_{\pm}(l)] =
                        e^{\pm \complexi \varphi}\left(\pm \cos{\theta}\frac{\partial}{\partial\theta}
                        + \frac{\complexi}{\sin{\theta}}\frac{\partial}{\partial\varphi} - l \sin{\theta} \right) \,, \\
  \operator{A}^{\pm}_{\mp}(l) &\coloneqq \mp [\operator{L}_{\pm}, \operator{J}_{\mp}(l)] =
                        e^{\pm \complexi \varphi}\left(\pm \cos{\theta}\frac{\partial}{\partial\theta}
                        + \frac{\complexi}{\sin{\theta}}\frac{\partial}{\partial\varphi} + l \sin{\theta} \right) \,.
\end{align}
The square brackets denote the commutator of two operators, i.e.
$[\operator{A}, \operator{B}]\coloneqq \operator{A}\operator{B} - \operator{B}\operator{A}$. These
operators shift $l$ and $m$ either in the same ``direction'' or in opposite ``directions''. Their
action is described by \citet[eq. 26 a,b, eq. 35 a,b]{Fakhri2016} and it is
\begin{equation}
  \begin{split}
    \operator{A}^{+}_{+}(l + 1)Y^{m}_{l} &= \sqrt{\frac{2l + 1}{2l + 3} (l + m + 2) (l + m + 1)} Y^{m + 1}_{l+1} \\
    \operator{A}^{-}_{-}(l)Y^{m}_{l} &= \sqrt{\frac{2l + 1}{2l - 1} (l + m) (l + m - 1)} Y^{m - 1}_{l - 1} \\
    \operator{A}^{+}_{-}(l)Y^{m}_{l} &= \sqrt{\frac{2l + 1}{2l - 1} (l - m) (l - m - 1)} Y^{m + 1}_{l-1} \\
    \operator{A}^{-}_{+}(l + 1)Y^{m}_{l} &= \sqrt{\frac{2l + 1}{2l + 3} (l - m + 2) (l - m + 1)} Y^{m - 1}_{l + 1} \,.
  \end{split}
\end{equation}
The operators $\operator{A}^{y}$ and $\operator{A}^{z}$ can be represented with the help of the
above ladder operators, i.e.
\begin{align}
  \label{eq:Ay-shift-operators}
  \operator{A}^{y} &= \frac{1}{2(2l + 1)} \left(-\operator{A}^{+}_{+}(l + 1) + \operator{A}^{+}_{-}(l)
                + \operator{A}^{-}_{+}(l) - \operator{A}^{-}_{-}(l + 1)\right) \\
  \label{eq:Az-shift-operators}
  \operator{A}^{z} &= \frac{i}{2(2l + 1)} \left(\operator{A}^{+}_{+}(l + 1) - \operator{A}^{+}_{-}(l)
                + \operator{A}^{-}_{+}(l) - \operator{A}^{-}_{-}(l + 1)\right) \,.
\end{align}
This can be verified by direct computation.

Knowing the action of the direction operators (see eq.~\eqref{eq:recurrence_relation_xY},
\eqref{eq:action-Ay} and \eqref{eq:action-Az}) , we can now compute their matrix representations.
This yields
\begin{align}
  \label{eq:Ax}
  (\mathbfss{A}^{x})_{i(l',m')j(l,m)}
  &= \sqrt{\frac{(l + m + 1)(l - m + 1)}{(2l + 3)(2l + 1)}} \delta_{l'(l+1)}\delta_{m'm} + \sqrt{\frac{(l + m)(l - m)}{(2l + 1)(2l - 1)}} \delta_{l'(l-1)}\delta_{m'm}  \\
  \label{eq:Ay}
  (\mathbfss{A}^{y})_{i(l',m')j(l,m)} &= \frac{1}{2}\left(- a^{m+1}_{l+1}\delta_{l'(l+1)}\delta_{m'(m+1)} +  a^{-m}_{l}\delta_{l'(l-1)}\delta_{m'(m+1)}
                                      +  a^{-m+1}_{l+1}\delta_{l'(l+1)}\delta_{m'(m-1)} - a^{m}_{l}\delta_{l'(l-1)}\delta_{m'(m-1)}\right)  \\
  \label{eq:Az}
  (\mathbfss{A}^{z})_{i(l',m')j(l,m)} & = \frac{\complexi}{2}\left( + a^{m+1}_{l+1}\delta_{l'(l+1)}\delta_{m'(m+1)} -  a^{-m}_{l} \delta_{l'(l-1)} \delta_{m'(m + 1)}
                                        +  a^{-m+1}_{l+1}\delta_{l'(l+1)}\delta_{m'(m-1)} - a^{m}_{l} \delta_{l'(l-1)}\delta_{m'(m-1)}\right)
\end{align}
These matrix elements can also be found in \citet[][Chapter 13.2, eq.
14-16]{Varshalovich_QuantumTheoryOfAngularMomentum}.

\subsection{Products of operators}
\label{sec:nabla_p-in-terms-of-operators}

The only term in the VFP equation that cannot immediately be formulated using the identity, angular momentum and direction operators is
\begin{equation}
  \label{eq:fictitious-force}
   -\left(\gamma m \frac{\mathrm{d} \mathbfit{U}}{\mathrm{d} t}
    + (\mathbfit{p} \cdot \nabla_{x}) \mathbfit{U} \right) \cdot \nabla_{p} f \, .
\end{equation}
This term is a consequence of the transformation of $\mathbfit{p}$ to the local fluid frame and 
is referred to as a \textit{fictitious force term} \citep{Achterberg2018_1}.

We show now that $\nabla_{p}$ can be expressed in terms of the
direction operators $\operator{A}^{a}$ and the angular momentum operators $\operator{L}^{a}$:
\begin{align}
  \label{eq:nabla-p-spherical-coordinates}
  \nabla_{p} = \mathbfit{e}_{p}\frac{\partial}{\partial p} + \frac{1}{p}\left(\mathbfit{e}_{\theta} \frac{\partial}{\partial \theta} + \mathbfit{e}_{\varphi}\frac{1}{\sin\theta} \frac{\partial}{\partial \varphi}\right)
  \coloneqq \mathbfit{e}_{p}\frac{\partial}{\partial p} - \complexi \frac{1}{p} \mathbfit{e}_{p}\times \mathbfit{L} \,.
\end{align}
Here,  $\mathbfit{e}_{p}, \mathbfit{e}_{\theta}$ and $\mathbfit{e}_{\varphi}$, are unit vectors of the
spherical coordinate system, and we point out that the components of \(\mathbfit{e}_{p}\) are
just the direction operators:
\begin{align}
  \label{eq:nabla-p-momentum-part-A}
  \mathbfit{e}_{p} = (\cos\theta, \sin\theta \cos\varphi, \sin\theta \sin\varphi)^{T}
  = (\operator{A}^{x}, \operator{A}^{y}, \operator{A}^{z})^{T}\, .
\end{align}

Note that expressing $\nabla_{p}$ with the help of the direction operators and the angular
momentum operators comes at the cost of introducing products of operators. 
Up to now, we saw that it was possible to derive the system of
equations for the expansion coefficients by simply replacing the operators with their respective
matrix representations. We now prove that this is still true for products of operators if some care
is taken.

First of all, we show that the matrix representation of a product of operators is the product of the
matrix representations of the operators. To see this, we point out that the identity operator in the
space of the spherical harmonics can be written as
\begin{equation}
  \label{eq:unit-operator}
  g = \operator{1}g = \sum^{\infty}_{n = 0} \sum^{n}_{k = -n} Y^{k}_{n}\scalarproduct{Y^{k}_{n}}{g} \qquad \text{for all } g \in \mathcal{S} \,.
\end{equation}
If $g \in \mathcal{S}$, then it can be written as a linear combination of spherical harmonics and
the scalar products in the above sum yield the coefficients of this linear combination.\footnote{A more
familiar notation for the identity operator might be the Bra-Ket notation of quantum mechanics. In
it the identity operator is expressed as $\operator{1} = \sum_{n,k} |nk\!\!><\!\!nk|$. But that is just
another way to say the same thing.}

For the matrix representation of the product of two operators, say $\operator{A}$ and
$\operator{B}$, we get
\begin{equation}
  \label{eq:matrix-represenation-of-product-of-operators}
  \begin{split}
    \scalarproduct{Y^{m'}_{l'}}{\operator{A} \operator{B}Y^{m}_{l}}
    &= \scalarproduct{Y^{m'}_{l'}}{\operator{A} \operator{1} \operator{B}Y^{m}_{l}} \\
    &= \scalarproduct{Y^{m'}_{l'}}{\operator{A} \sum^{\infty}_{n = 0} \sum^{n}_{k = -n}  Y^{k}_{n} \scalarproduct{Y^{k}_{n}}{\operator{B}Y^{m}_{l}}} \\
    &= \sum^{\infty}_{n = 0} \sum^{n}_{k = -n} \scalarproduct{Y^{m'}_{l'}}{\operator{A}Y^{k}_{n} }\scalarproduct{Y^{k}_{n}}{\operator{B}Y^{m}_{l}} \\
    &= \sum^{\infty}_{n = 0} \sum^{n}_{k = -n} (\mathbfss{A})_{i(l',m')h(n,k)} (\mathbfss{B})_{h(n,k)j(l,m)} \,.
  \end{split}
\end{equation}
Hence, the matrix representations of the two operators are multiplied.

We emphasise that the above statement does \emph{not} hold if we truncate the expansion of
the distribution function at a finite $l_{\text{max}}$. In this case, we compute the matrix
representations of the operators acting on the finite dimensional space
$\mathcal{S}^{l_{\text{max}}} \coloneqq \text{span}\{Y^{0}_{0}, Y^{1}_{1}, Y^{0}_{1}, Y^{-1}_{1},
\dots, Y^{-l_{\text{max}}}_{l_{\text{max}}}\}$ and \emph{not} on the infinite dimensional space
$\mathcal{S}$. The $\operator{A}^{a}$ operators contain ladder operators which increase (or
decrease) $l$, see eq.~\eqref{eq:Ax-shift-operators},~\eqref{eq:Ay-shift-operators} and
eq.~\eqref{eq:Az-shift-operators}. It is necessary to define what happens if these operators act on
$Y^{m}_{l_{\text{max}}}$. Since $Y^{m}_{l_{\text{max} + 1}}$ is not an element of the space
$\mathcal{S}^{l_{\text{max}}}$, it makes sense to define that the result is zero when they act on
$Y^{m}_{l_{\text{max}}}$. But if a product of operators appears whose first operator increases $l$, while
the second operator decreases it again, then the above definition would imply that this product
yields zero when it acts on $Y^{m}_{l_{\text{max}}}$. This is not the correct result, because the
joint action results in a constant times $Y^{m}_{l_{\text{max}}}$, which can be represented in
$\mathcal{S}^{l_{\text{max}}}$. This difficulty can be avoided, if we compute the matrix
representations of the operators for $L = l_{\text{max}} + 1$, evaluate the products and,
subsequently, reduce the matrices to $l_{\text{max}}$.

Since we now know how to compute the matrix representations of a product of operators, we can now
look at eq.~\eqref{eq:vfp-mixed-spherical-exp} and see how the fictitious force term contributes to
the system of equations, i.e.
\begin{equation}
  \label{eq:fictiuous-force-terms}
  \begin{split}
    &- \int_{S^{2}} {Y^{m'}_{l'}}^{*} \left(\gamma m \frac{\mathrm{d} \mathbfit{U}}{\mathrm{d} t}
      + (\mathbfit{p} \cdot \nabla_{x}) \mathbfit{U} \right) \cdot \nabla_{p} (f^{m}_{l}Y^{m}_{l})\, \mathrm{d}\Omega = - \int_{S^{2}} {Y^{m'}_{l'}}^{*} \left(\gamma m \frac{\mathrm{d} \mathbfit{U}}{\mathrm{d} t}
      + (\mathbfit{p} \cdot \nabla_{x}) \mathbfit{U} \right) \cdot \left(\mathbfit{e}_{p}\partial_{p} - \frac{\complexi}{p} \mathbfit{e}_{p} \times \mathbfit{L}\right) f^{m}_{l}Y^{m}_{l}\, \mathrm{d}\Omega  \\
    &\qquad =   - \gamma m \frac{\mathrm{d} \mathbfit{U}}{\mathrm{d} t}  \cdot  \scalarproduct{Y^{m'}_{l'}}{
      \mathbfit{e}_{p}Y^{m}_{l}}\frac{\partial f^{m}_{l}}{\partial p}
      + \frac{\complexi}{V} \frac{\mathrm{d} \mathbfit{U}}{\mathrm{d} t}  \cdot \scalarproduct{Y^{m'}_{l'}}{\mathbfit{e}_{p} \times \mathbfit{L}Y^{m}_{l}}f^{m}_{l} \\
    &\qquad\phantom{=}
      \, \,{} - p\scalarproduct{Y^{m'}_{l'}}{(\mathbfit{e}_{p} \cdot \nabla_{x}) \mathbfit{U}  \cdot \mathbfit{e}_{p}Y^{m}_{l}} \frac{\partial f^{m}_{l}}{\partial p} + \complexi \scalarproduct{Y^{m'}_{l'}}{(\mathbfit{e}_{p} \cdot \nabla_{x}) \mathbfit{U}  \cdot (\mathbfit{e}_{p}\times \mathbfit{L})Y^{m}_{l}} f^{m}_{l} \\
    &\qquad = - \gamma m \frac{\mathrm{d} U_{a}}{\mathrm{d} t} \scalarproduct{Y^{m'}_{l'}}{
      \operator{A}^{a}Y^{m}_{l}}\frac{\partial f^{m}_{l}}{\partial p}
      +  \frac{\complexi}{V} \epsilon_{abc} \frac{\mathrm{d} U^{a}}{\mathrm{d} t} \scalarproduct{Y^{m'}_{l'}}{\operator{A}^{b}\operator{L}^{c}Y^{m}_{l}}f^{m}_{l} \\
    &\qquad\phantom{=}
      \, \,{} - p \frac{\partial U_{b}}{\partial x^{a}} \scalarproduct{Y^{m'}_{l'}}{\operator{A}^{a} \operator{A}^{b}Y^{m}_{l}}\frac{\partial f^{m}_{l}}{\partial p} + \complexi \epsilon_{bcd} \frac{\partial U_{b}}{\partial x^{a}} \scalarproduct{Y^{m'}_{l'}}{\operator{A}^{a}  \operator{A}^{c}\operator{L}^{d}Y^{m}_{l}} f^{m}_{l} \\
    &\qquad = - \gamma m \frac{\mathrm{d} U_{a}}{\mathrm{d} t}  
      (\mathbfss{A}^{a})_{i(l',m')j(l,m)} \partial_{p} \mathbfit{f}_{j(l,m)}
      +  \sum_{n,k} \frac{\complexi}{V} \epsilon_{abc} \frac{\mathrm{d} U^{a}}{\mathrm{d} t} (\mathbfss{A}^{b})_{i(l',m')h(n,k)}(\bm{\Omega}^{c})_{h(n,k)j(l,m)} \mathbfit{f}_{j(l,m)} \\
    &\qquad\phantom{=}
      - \sum_{n,k} p \frac{\partial U_{b}}{\partial x^{a}} (\mathbfss{A})^{a}_{i(l',m')h(n,k)} (\mathbfss{A})^{b}_{h(n,k)j(l,m)}\partial_{p} \mathbfit{f}_{j(l,m)} \\
    &\qquad\phantom{=} 
      \, \,{} + \sum_{n,k} \sum_{n',k'} \complexi \epsilon_{bcd} \frac{\partial U_{b}}{\partial x^{a}} (\mathbfss{A})^{a}_{i(l',m')h(n,k)}  (\mathbfss{A})^{c}_{h(n,k)r(n',k')}(\bm{\Omega})^{d}_{r(n',k')j(l,m)} \mathbfit{f}_{j(l,m)} \,.
  \end{split}
\end{equation}

We stress that it is exactly this computation where the proposed operator method provides the most obvious benefit. 
The repeated application of recurrence relations for the associated Legendre polynomials is
replaced with the evaluation of matrix products, which avoids lengthy and cumbersome calculations \cite[see][Appendix A]{Reville2013}. 

\subsection{The complete system of equations}
\label{sec:complete-system-of-equations}

We conclude this section with a presentation of the complete system of equations, which determines
the expansion coefficients $f^{m}_{l}$, and an explanation of its terms. The system of equations is
\begin{equation}
  \label{eq:complex-system-of-equations}
  \begin{split}
    &\left(\mathbfss{1} +  \frac{V}{c^{2}} U_{a} \mathbfss{A}^{a}\right) \partial_{t}\mathbfit{f}
    + \left(U^{a}\mathbfss{1} + V \mathbfss{A}^{a}\right) \partial_{x_{a}}\mathbfit{f}
    - \left(\gamma m \frac{\mathrm{d} U_{a}}{\mathrm{d} t} \mathbfss{A}^{a}
    + p \frac{\partial U_{b}}{\partial x^{a}} \mathbfss{A}^{a}\mathbfss{A}^{b}\right)\partial_{p} \mathbfit{f} \\
    &\qquad {} + \left(\frac{\complexi}{V} \epsilon_{abc} \frac{\mathrm{d} U^{a}}{\mathrm{d} t} \mathbfss{A}^{b}\bm{\Omega}^{c}
    + \complexi \epsilon_{bcd} \frac{\partial U_{b}}{\partial x^{a}}\mathbfss{A}^{a}\mathbfss{A}^{c}\bm{\Omega}^{d}\right) \mathbfit{f}
    - \complexi \omega_{a}\bm{\Omega}^{a}\mathbfit{f} + \nu \mathbfss{C}\mathbfit{f}
    =  0\,.
  \end{split}
\end{equation}

The first term is the time derivative of the expansion coefficients including a correction of time
stemming from the transformation of momentum variables into the rest frame of the magnetised (or background) plasma.
Its second term models how the expansion coefficients change because of the motion of the plasma,
which is described by $\mathbfit{U}$, and the motion of the particles given through
$V\mathbfss{A}^{a}$. As already stated above, we refer to this term as the spatial advection term.
The next term changes the energy contained in the expansion coefficients because of the fictitious
force acting on the particles. We call it the \textit{momentum advection term}, because it advects
the coefficients in $p$-direction. The fourth term reflects that the fictitious force not only
accelerates (or decelerates), but it also changes how the energy is distributed among the expansion
coefficients. The fifth term represents the effect of the magnetic force, namely the rotation of the
velocity of the particles. This rotation translates into a rotation of the expansion
coefficients.\footnote{For readers familiar with rotations in the space of spherical harmonics (or
  the theory of angular momentum in quantum mechanics): It can be shown, that
  $\partial_{t} f(\mathbfit{x},\mathbfit{p},t) - \bomega \cdot (\mathbfit{p} \times \nabla_{p})
  f(\mathbfit{x},\mathbfit{p},t) = 0$ has the solution
  $f(\mathbfit{x}, \mathbfit{p}, t) = e^{i t \bomega \cdot \mathbfit{L}}
  f(\mathbfit{x},\mathbfit{p}, 0) = f(\mathbfit{x},\mathbfss{R}(t) \mathbfit{p}, 0)$.
  Here, $\mathbfss{R}(t)$ is a rotation matrix, which rotates $\mathbfit{p}$ about $\bomega/\omega$ at the
  angular frequency $\omega$. We express $f$ as a series of spherical harmonics and rotating
  spherical harmonics is equivalent to expressing a spherical harmonic in terms of other spherical
  harmonics of the same degree $l$. This is what the $\bm{\Omega}^{a}$ matrices, which are
  representation matrices of the angular momentum operator, do.} We refer to this term as the
\textit{rotational term}. The last term is modelling the interactions of the particles with the
thermal plasma. $\mathbfss{C}$ is a diagonal matrix, which causes the expansion coefficients with
$l \geq 1$ to decay exponentially at the rate $\nu l(l+1)/2$. The coefficients with $l \geq 1$ model
the anisotropies of the distribution function and the effect of the exponential decay is to drive
the distribution towards isotropy. We call this term \textit{collision term}.

We, once more, would like to highlight the one-to-one correspondence between operators and
representation matrices. The system of equations can be derived by simply replacing the operators
with their corresponding matrices.

\section{Rotations in the spherical harmonic space}
\label{sec:rotations-spherical-harmonics}

In this section we present an alternative way to compute the representation matrices
$\mathbfss{A}^{y}, \mathbfss{A}^{z}, \bm{\Omega}^{y}$ and $\bm{\Omega}^{z}$. It is based on the
observation that our choice of the coordinate system in momentum space was arbitrary and that the
direction operators $\operator{A}^{a}$ and the angular momentum operators $\operator{L}^{a}$ are
defined with respect to the coordinate axes, see their definitions in eq.~\eqref{eq:A-operators}
and eq.~\eqref{eq:angular-momentum-operators} respectively. The spherical harmonics are also defined in the same coordinate system. Rotating them clockwise by $\upi/2$ about the $x$-axis is equivalent to defining them in an equally rotated coordinate system. In the rotated system the former $y$-axis is the $z$-axis and we expect the representation matrix $\mathbfss{A}_{y}$ to have transformed to $\mathbfss{A}_{z}$. Taking advantage of this requires us to investigate how to rotate the spherical harmonics and how
the operators and their representation matrices change under such a rotation.

\subsection{Deriving the rotation operator}
\label{sec:rotation-operator}

We begin our investigation with formally introducing a rotation operator in the space of spherical
harmonics, i.e. $Y^{\tilde{m}}_{l}(\theta, \varphi) = \operator{R}Y^{m}_{l}(\theta, \varphi)$.
$\tilde{m}$ reflects the idea that we can think of the rotated spherical harmonics as being defined
with respect to a new polar direction given by a corresponding rotation of the original one. A
direct consequence of this thought is that a rotation does not change the degree $l$ of the
spherical harmonic.

An explicit form of the rotation operator $\operator{R}$ can be derived by defining its action on
the spherical harmonics and by studying an infinitesimal rotation. This is, in essence, an
application of representation theory. We present the details here for completeness. In doing so, we follow
\citet{Jeevanjee_TensorsGroups}, in particular Chapter 5 and example 5.9 therein.

We point out that
$Y^{m}_{l}(\theta ,\varphi) = N^{m}_{l} P^{m}_{l}(\cos\theta) e^{im\varphi} = N^{m}_{l}
P^{m}_{l}(p^{x}/p) e^{i m \arctan{(p^{z}/p^{y}})} = Y^{m}_{l}(\mathbfit{p})$ and define the action
of $\operator{R}$ to be
\begin{equation}
  \label{eq:action-rotation-operator}
  Y^{\tilde{m}}_{l}(\mathbfit{p}) = \operator{R} Y^{m}_{l}(\mathbfit{p}) \coloneqq
  Y^{m}_{l}(\mathbfss{R}^{-1}\mathbfit{p}) \,.
\end{equation}
Where $\mathbfss{R}$ is a rotation matrix which rotates the vector $\mathbfit{p}$.

We now look at an infinitesimal rotation to derive an explicit expression for the rotation operator.
To simplify the discussion, we restrict ourselves to a rotation about the $x$-axis by an
infinitesimally small angle $\mathrm{d}\alpha$, i.e.
$\mathbfss{R}_{x}(\mathrm{d}\alpha) \mathbfit{p} = \mathbfit{p} + \mathrm{d}\alpha \mathbfit{e}_{x}
\times \mathbfit{p}$. Its inverse is
$\mathbfss{R}^{-1}_{x}(\mathrm{d}\alpha) = \mathbfit{p} - \mathrm{d}\alpha \mathbfit{e}_{x} \times
\mathbfit{p}$.

An explicit expression for the rotation operator is now constructed by plugging the infinitesimal
rotation into the definition of its action~\eqref{eq:action-rotation-operator}. This yields
\begin{equation}
  \label{eq:rotation-operator-spherical-harmonics}
  \begin{split}
    Y^{\tilde{m}}_{l}(\mathbfit{p}) = Y^{m}_{l}(\mathbfit{p} - \mathrm{d}\alpha
    \mathbfit{e}_{x} \times \mathbfit{p})
    &= Y^{m}_{l}(\mathbfit{p}) -
    \frac{\partial Y^{m}_{l}}{\partial p^{a}}\Bigg|_{\mathrm{d} \alpha = 0}
    \epsilon_{abc}(\mathbfit{e}_{x})^{b} (\mathbfit{p})^{c} \mathrm{d}\alpha +
    \mathcal{O}(\mathrm{d}\alpha^{2}) \\
    &= Y^{m}_{l}(\mathbfit{p}) - (\mathbfit{p}
    \times \nabla_{p})_{x} Y^{m}_{l}(\mathbfit{p}) \mathrm{d}\alpha +
      \mathcal{O}(\mathrm{d}\alpha^{2}) \\
    &= \left(1 - \complexi \mathrm{d}\alpha \operator{L}^{x} \right) Y^{m}_{l}(\mathbfit{p}) +
    \mathcal{O}(\mathrm{d}\alpha^{2}) \,.
  \end{split}
\end{equation}
Where we Taylor expanded in $\mathrm{d}\alpha$ at zero.

In a next step, we set $\mathrm{d}\alpha = \alpha/n $ and perform $n$ infinitesimal rotations, which
add up to a rotation about the $x$-axis by an angle $\alpha$. In the limit $n \rightarrow \infty$,
we get
\begin{equation}
  \label{eq:rotation-operator-exponential-form}
  Y^{\tilde{m}}_{l}(\mathbfit{p}) = \lim_{n \rightarrow \infty}\left(1 - \complexi \frac{\alpha}{n} \operator{L}^{x}\right)^{n} Y^{m}_{l}(\mathbfit{p}) = e^{-\complexi \alpha\operator{L}^{x}}Y^{m}_{l}(\mathbfit{p}) \,.
\end{equation}

We state, without further proof, that this generalises to
$\operator{R} = e^{-\complexi \alpha \mathbfit{r} \cdot \mathbfit{L}}$ for an arbitrary rotation
about the \emph{unit} vector $\mathbfit{r}$ by an angle $\alpha$.

The matrix elements of the rotation operator $\operator{R}$ 
are the \textit{Wigner-D functions}, which are well-known
in the quantum theory of angular momentum, see \citet[Chapter 4.5, eq.
1]{Varshalovich_QuantumTheoryOfAngularMomentum}. They are defined to be
\begin{equation}
  \label{eq:wigner-d-functions}
  \begin{split}
    (\mathbfss{U})_{i(l',m')j(l,m)}(\alpha \mathbfit{r})
    \coloneqq \scalarproduct{Y^{m'}_{l'}}{e^{-\complexi \alpha \mathbfit{r} \cdot \mathbfit{L}} Y^{m}_{l}}
    &= \sum^{\infty}_{k = 0} \frac{1}{k!}\scalarproduct{Y^{m'}_{l'}}{\left(-\complexi \alpha \mathbfit{r} \cdot \mathbfit{L}\right)^{k} Y^{m}_{l}} \\
    &= \sum^{\infty}_{k = 0} \frac{1}{k!} (-\complexi \alpha r_{a} \bm{\Omega}^{a})^{k}_{i(l',m')j(l,m)}
    = (e^{-\complexi \alpha r_{a} \bm{\Omega}^{a}})_{i(l',m')j(l,m)} \,.
  \end{split}
\end{equation}
The conjugate transpose of the Wigner-D matrix is
\begin{equation}
  \label{eq:conjugate-transpose-wigner-d}
  \mathbfss{U}^{\bm{\dagger}} = (e^{-\complexi \alpha r_{a} \bm{\Omega}^{a}})^{\bm{\dagger}}
  =\sum^{\infty}_{k = 0} \frac{1}{k!}(-\complexi \alpha r_{a}\bm{\Omega}^{a})^{k\bm{\dagger}}
  =\sum^{\infty}_{k = 0} \frac{1}{k!}(\complexi \alpha r_{a}\bm{\Omega}^{a\bm{\dagger}})^{k}
  =\sum^{\infty}_{k = 0} \frac{1}{k!}(\complexi \alpha r_{a}\bm{\Omega}^{a})^{k}
  =e^{\complexi \alpha r_{a} \bm{\Omega}^{a}} \,,
\end{equation}
where we used that for arbitrary matrices $\mathbfss{X}$ and $\mathbfss{Y}$ the following two
properties hold: Firstly, for
$(\mathbfss{X} \cdots \mathbfss{X})^{\bm{\dagger}} = \mathbfss{X}^{\bm{\dagger}} \cdots
\mathbfss{X}^{\bm{\dagger}}$ the order of the matrices does not matter and, second, that
$(\mathbfss{X} + \mathbfss{Y})^{\bm{\dagger}} = \mathbfss{X}^{\bm{\dagger}} +
\mathbfss{Y}^{\bm{\dagger}}$. Moreover, we took advantage of the fact that the representation
matrices of the angular momentum operator are Hermitian, namely
$\bm{\Omega}^{a\bm{\dagger}} = \bm{\Omega}^{a}$, see eq.~(\ref{eq:Omega-x-hermitian},
\ref{eq:Omega-y-hermitian}).

A direct and useful consequence is that the Wigner-D matrix is unitary\footnote{In quantum mechanics
  the unitarity of the rotation operator implies that the probability to measure a specific value of
  the angular momentum does not change if the coordinate system is rotated.}, i.e.
\begin{equation}
  \label{eq:Wigner-D-matrix-unitary}
  \mathbfss{U}^{\bm{\dagger}}\mathbfss{U} = e^{\complexi \alpha r_{a} \bm{\Omega}^{a}} e^{-\complexi \alpha r_{a} \bm{\Omega}^{a}} = \mathbfss{1} \,.
\end{equation}

\subsection{The change of operators and their representation matrices under a rotation of the
  coordinate system}
\label{sec:change-operator-rotation}

Knowing how to rotate the spherical harmonics, we are in a position to compute the action of an
operator $\operator{O}$ in the rotated coordinate frame.

We once more note that we can compute the spherical harmonics defined in the rotated frame simply by rotating
the original spherical harmonics in the same manner as the coordinate system.

Before we derive the action of an operator in the rotated coordinate system, we would like to make a
technical remark: Transformations of coordinates can be interpreted \textit{actively} or
\textit{passively}. An active rotation, for example, yields the coordinates of a rotated vector,
whereas a passive rotation gives the coordinates of the vector in a rotated coordinate frame. Active
and passive rotations are equivalent if the sign of the rotation angle is changed, i.e.
$\mathbfss{R}^{\rm pas}(\alpha \mathbfit{r}) = \mathbfss{R}^{\rm act}(-\alpha \mathbfit{r})$. Even though we
write as if we rotated the coordinate system, we actually interpret the rotations as \emph{active}.
Instead of working with two different coordinate systems in which the spherical harmonics are
defined, we work with two different bases of the space of spherical harmonics, namely the original
spherical harmonics and the rotated spherical harmonics. Both are defined in the original coordinate
system. The reason to take the active point of view is to allow for a direct application of
representation theory without the difficulties from two different sets of coordinates, for example,
$\tilde{\mathbfit{p}}$ in the rotated and $\mathbfit{p}$ in the original coordinate frame.

That said, let $\tilde{O}$ be the operator in the rotated coordinate system, then
$\tilde{O}Y^{\tilde{m}}_{l}(\theta, \varphi) =
\operator{R}\operator{O}\operator{R}^{-1}Y^{\tilde{m}}_{l}(\theta, \varphi)$; the inverse rotation
$\operator{R}^{-1}$ ``brings'' the spherical harmonic back into the unrotated coordinate system. The
action of $\operator{O}$ on the unrotated spherical harmonic is known. Its result is rotated again
and $\tilde{O}$ gives, as expected, results in the rotated coordinate system.

Eventually, we are interested in the representation matrices of operators and, thus, we would like
to know how the representation matrix of an operator $\operator{O}$ changes under a rotation of the
coordinate system. The representation matrix $\tilde{\mathbfss{O}}$ in the rotated frame is given by
the action of $\operator{O}$ on the rotated spherical harmonics, i.e.
\begin{equation}
  \label{eq:change-representation-matrix-rotation}
  \begin{split}
    (\tilde{\mathbfss{O}})_{i(l',m')j(l,m)} = \scalarproduct{Y^{\tilde{m}'}_{l'}}{\operator{O}Y^{\tilde{m}}_{l}}
    &= \scalarproduct{\operator{R} Y^{m'}_{l'}}{ \operator{O} \operator{R} Y^{m}_{l}} \\
    &= \sum_{n,k}\sum_{n',k'} \scalarproduct{\operator{R} Y^{m'}_{l'}}{Y^{k'}_{n'}} \scalarproduct{Y^{k'}_{n'}}{\operator{O} Y^{k}_{n}} \scalarproduct{Y^{k}_{n}}{\operator{R} Y^{m}_{l}} \\
    &= \sum_{n,k}\sum_{n',k'} (\mathbfss{U}^{\bm{\dagger}})_{i(l',m')h(n',k')} (\mathbfss{O})_{h(n',k')r(n,k)} (\mathbfss{U})_{r(n,k)j(l,m)} \,.
  \end{split}
\end{equation}
Or compactly, $\tilde{\mathbfss{O}} = \mathbfss{U}^{\bm{\dagger}} \mathbfss{O} \mathbfss{U}$.

\subsection{Computing the representation matrices using rotation matrices}
\label{sec:computing-representation-matrices-rotation-matrices}

As explained in the introduction of this section, the operator $\operator{A}^{y}$ has to be
$\operator{A}^{z}$ in a coordinate system which is rotated by $-\upi/2$ about the $x$-axis. If we know the representation matrix $\mathbfss{A}^{y}$,
we can compute how it looks like in the rotated frame using the formula in
eq.~\eqref{eq:change-representation-matrix-rotation} and we know that in this frame it has to equal
$\mathbfss{A}^{z}$. Hence,
\begin{equation}
  \label{eq:A-z-from-A-y}
    \mathbfss{A}^{z} =  e^{-i\frac{\upi}{2}\bm{\Omega}^{x}}\mathbfss{A}^{y}e^{i\frac{\upi}{2}\bm{\Omega}^{x}}
\end{equation}
Moreover, we compute the representation matrix $\mathbfss{A}^{y}$ knowing $\mathbfss{A}^{x}$ by
rotating the original coordinate system about the $z$-axis by $-\upi/2$, i.e.
\begin{equation}
  \label{eq:A-y-from-A-x}
  \mathbfss{A}^{y}= e^{-i\frac{\upi}{2}\bm{\Omega}^{z}}\mathbfss{A}^{x}e^{i\frac{\upi}{2}\bm{\Omega}^{z}}
\end{equation}

This leads us to the conclusion, that it is enough to compute the representation matrix of
$\operator{A}^{x}$. We highlight that we do \emph{not} have to compute the representation matrices
of the involved rotations ourselves. Since their matrix elements are the Wigner-D functions, they
can be, for example, be found in \citet[Section 4.5.6 eq. 29, eq. 30 and Section 4.3.1, eq.
2]{Varshalovich_QuantumTheoryOfAngularMomentum}. They are
\begin{equation}
  \label{eq:explicit-rotation-matrices}
  \begin{split}
    (\mathbfss{U})_{i(l',m')j(l,m)}(-\upi/2 \mathbfit{e}_{x}) &= e^{i\frac{\upi}{2}\bm{\Omega}^{x}}
                                                                = \delta_{l'l}\delta_{m'm}e^{im\upi/2} \quad \text{and}\\
    (\mathbfss{U})_{i(l',m')j(l,m)}(-\upi/2 \mathbfit{e}_{z}) &= e^{i\frac{\upi}{2}\bm{\Omega}^{z}}
                                                                = \delta_{l'l}  \frac{(-1)^{l - m'}}{2^{l}} \sum^{n}_{\substack{k = 0\\ m' + m + k \geq 0}} (-1)^{k}
                                                                \frac{\left[(l + m')!(l - m')!(l + m)!(l - m)!\right]^{1/2}}{k!(l - m' - k)!(l - m - k)!(m' + m + k)!}
  \end{split} \,,
\end{equation}
where the limit of the sum is $n = \min(l-m', l-m)$.

We remark that the representation matrices of the angular momentum operators can as well be obtained
by rotating $\bm{\Omega}^{x}$, i.e.
\begin{equation}
  \label{eq:Omega-y-and-Omega-z-from-Omega-x}
    \bm{\Omega}^{y}  = e^{-i\frac{\upi}{2}\bm{\Omega}^{z}}\bm{\Omega}^{x}e^{i\frac{\upi}{2}\bm{\Omega}^{z}} \quad\text{and}\quad
    \bm{\Omega}^{z}  = e^{-i\frac{\upi}{2}\bm{\Omega}^{x}}\bm{\Omega}^{y}e^{i\frac{\upi}{2}\bm{\Omega}^{x}} \,.
\end{equation}

We emphasise that to construct the system of equations~\eqref{eq:complex-system-of-equations}, we
only need four matrices, namely
$\mathbfss{A}^{x}, \bm{\Omega}^{x}, \mathbfss{U}(-\upi/2 \mathbfit{e}_{x})$, and
$\mathbfss{U}(-\upi/2 \mathbfit{e}_{z})$ and two of them are diagonal matrices. This should
considerably diminish the burden of its implementation in future numerical applications.

\section{A real system of equations}
\label{sec:real-system-of-equations}

The phase-space density $f$ is a real
function, i.e $f \in \mathbb{R}$. This implies the existence of a relation between the coefficients
$f^{m}_{l}$ of the spherical harmonic expansion of $f$ as given in
eq.~\eqref{eq:spherical-harmonic-exp}, because the results of the sums must yield a real function.
Since a scalar (function) $f$ is real if and only if $f^{*} = f$, the relation between the
coefficients can be shown to be
\begin{equation}
  \label{eq:complex-and-real-expansion-coefficients}
  f^{m}_{l} = \scalarproduct{Y^{m}_{l}}{f} = \scalarproduct{Y^{m}_{l}}{f^{*}}
  = (-1)^{m}{f^{-m}_{l}}^{*} \,,  
\end{equation}
where we used that ${Y^{m}_{l}}^{*} = (-1)^{m} Y^{-m}_{l}$. This means that the system of
equations~\eqref{eq:complex-system-of-equations} contains too much information, i.e. too many
equations, because it determines the expansion coefficients $f^{-m}_{l}$ as well. And these, as we
have just proven, can be directly obtained from the ones with positive $m$. A way to remedy this is
to use real spherical harmonics instead of complex spherical harmonics. We note that the above derivations are greatly simplified by working with the complex spherical harmonics, and this is why we choose to transform to the real spherical harmonics only after the matrix representations have been determined. 

\subsection{Definition of the real spherical harmonics and the basis transformation}
\label{sec:basis-transformation}

We begin with a definition of the \textit{real spherical harmonics}. One way to arrive at their
definition is to rewrite the spherical harmonic expansion~\eqref{eq:spherical-harmonic-exp} as
\begin{equation}
  \label{eq:real-spherical-harmonic-expansion}
  \begin{split}
    f &= \sum^{\infty}_{l = 0} \sum^{l}_{m = -l} f^{m}_{l}Y^{m}_{l}
    = \sum^{\infty}_{l = 0} f^{0}_{l} Y^{0}_{l} + \sum^{l}_{m = 1} \left(f^{m}_{l}Y^{m}_{l} + f^{-m}_{l}Y^{-m}_{l}\right) 
    = \sum^{\infty}_{l = 0} f^{0}_{l} Y^{0}_{l} + \sum^{l}_{m = 1} \left(f^{m}_{l}Y^{m}_{l} + {f^{m}_{l}}^{*}{Y^{m}_{l}}^{*}\right) \\
      &= \sum^{\infty}_{l = 0} f^{0}_{l} Y^{0}_{l}
        + \sum^{l}_{m = 1} \left(\frac{f^{m}_{l} + {f^{m}_{l}}^{*}}{\sqrt{2}} \frac{Y^{m}_{l} + {Y^{m}_{l}}^{*}}{\sqrt{2}}
        - \frac{(f^{m}_{l} - {f^{m}_{l}}^{*})}{\sqrt{2}\complexi} \frac{(Y^{m}_{l} - {Y^{m}_{l}}^{*})}{\sqrt{2}\complexi}\right) \\
    &\coloneqq \sum^{\infty}_{l = 0} f_{l00} Y_{l00}
      + \sum^{l}_{m = 1} \left( f_{lm0} Y_{lm0} + f_{lm1} Y_{lm1} \right)
      = \sum^{\infty}_{l = 0} \sum^{l}_{m = 0} \sum^{1}_{s =0 } f_{lms} Y_{lms}
      = \sum^{\infty}_{l = 0} \sum^{1}_{s =0 } \sum^{l}_{m = \vphantom{0}s}  f_{lms} Y_{lms} \, .
  \end{split}
\end{equation}
In the last line we defined the real spherical harmonics $Y_{lms}$. 

The expansion coefficients $f_{lms}$ are real as well. Furthermore, $f_{l01} = 0$ and $Y_{l01}= 0$.

The real spherical harmonics are orthogonal functions too, i.e.
$\scalarproduct{Y_{l'm's'}}{Y_{lms}} = \delta_{l'l} \delta_{m'm} \delta_{s's}$, which can be
verified through a direct computation of the scalar product.

We can interpret the rewrite of eq.~\eqref{eq:spherical-harmonic-exp} as a basis transformation in
the space of spherical harmonics $\mathcal{S}$. The corresponding basis transformation
matrix\footnote{In eq. (5.1) \citet{Schween2022} we as well introduced a matrix $\mathbfss{S}$,
  which relates the real and complex spherical harmonics. We remark that it is the transpose of the
  basis transformation matrix we define here.} can be computed with the help of
\begin{equation}
  \label{eq:basis-transformation}
  Y_{l'm's'} = \sum^{\infty}_{l = 0} \sum^{l}_{m = -l} Y^{m}_{l} \scalarproduct{Y^{m}_{l}}{Y_{l'm's'}}
  \coloneqq \sum^{\infty}_{l = 0} \sum^{l}_{m = -l} Y^{m}_{l} (\mathbfss{S})_{i(l,m)j(l',m',s')} \, .
\end{equation}
We introduced a new index map $j(l',m',s')$. It again encodes the ordering of the real spherical
harmonics $Y_{lms}$ and their coefficients $f_{lms}$. For example, if the real spherical harmonics
are ordered as
$\mathcal{S}=\text{span}\{Y_{000}, Y_{110}, Y_{100}, Y_{111}, Y_{220}, Y_{210}, Y_{200}, Y_{211},
Y_{221}, \dots\}$, the index map is
$j(l',m',s') = l'(l'+1) + (-1)^{s'+1}m'$\footnote{\label{fn:no-m-zero-s-one}Note since there is no real
  spherical harmonic $Y_{l01}$, we define that the index map $j(l',m', s')$ is valid for $m'=0$ and
  $s'=0$ and \emph{exclude} $m'=0$ and $s'=1$.}. Note the index $j$ starts at zero.

The matrix elements of the basis transformation matrix $\mathbfss{S}$ can be computed to be
\begin{equation}
  \label{eq:basis-transformation-matrix-elements}
  (\mathbfss{S})_{i(l,m) j(l',m',s')} = \delta_{l'l}\left(
    \frac{1}{\sqrt{2(1 + \delta_{m'0})}}\left(\delta_{m'm} + (-1)^{m'}\delta_{-m'm}\right) \delta_{s'0}
    - \frac{\complexi}{\sqrt{2}}\left(\delta_{m'm} - (-1)^{m'}\delta_{-m'm}\right) \delta_{s'1} \right)
\end{equation}
where we used the definition of the real spherical harmonics given in
eq.~\eqref{eq:real-spherical-harmonic-expansion} and, once more, the relation
${Y^{m}_{l}}^{*} = (-1)^{m} Y^{-m}_{l}$.

Moreover, the basis transformation matrix has the useful property that it is a unitary matrix. This
is a direct consequence of the fact that basis transformation matrices relating two orthogonal bases
are unitary. We include a proof of this general statement for the case at hand, i.e.
\begin{equation}
  \label{eq:basis-transformation-unitary}
  \begin{split}
    (\mathbfss{1})_{i(l',m',s')j(l,m,s)} = \delta_{l'l}\delta_{m'm}\delta_{s's} = \scalarproduct{Y_{l'm's'}}{Y_{lms}}
    &= \sum_{n,k}\scalarproduct{Y^{k}_{n}}{Y_{l'm's'}}^{*} \scalarproduct{Y^{k}_{n}}{Y_{lms}}\\
    &= \sum_{n,k}(\mathbfss{S}^{\bm{\dagger}})_{i(l',m',s')h(n,k)} (\mathbfss{S})_{h(n,k)j(l,m,s)}\,.
  \end{split}
\end{equation}
In matrix form this equation reads $\mathbfss{1} = \mathbfss{S}^{\bm{\dagger}}\mathbfss{S}$ and the
inverse basis transformation matrix is $ \mathbfss{S}^{-1} = \mathbfss{S}^{\bm{\dagger}}$.

\subsection{Turning the complex system of equations into a real system of equations}
\label{sec:turning-complex-system-into-real-system}

With the basis transformation and its inverse at hand, we can turn the complex system of
equations~\eqref{eq:complex-system-of-equations} into a real system thereby removing the superfluous
equations. To this end, we compute how the matrix representation of an operator changes under the
basis transformation. We find for an arbitrary operator, say $\operator{O}$, that
\begin{equation}
  \label{eq:basis-transformation-matrix-representation}
  \begin{split}
    (\mathbfss{O}_{R})_{i(l',m',s')j(l,m,s)} &= \scalarproduct{Y_{l'm's'}}{\operator{O}Y_{lms}}
                                               = \sum_{n,k} \sum_{n',k'}  \scalarproduct{Y^{k}_{n}}{Y_{l'm's'}}^{*} \scalarproduct{Y^{k}_{n}}{\operator{O} Y^{k'}_{n'}} \scalarproduct{Y^{k'}_{n'}}{Y_{lms}} \\
        &= \sum_{n,k} \sum_{n',k'}  (\mathbfss{S}^{\bm{\dagger}})_{i(l',m',s')h(n,k)} (\mathbfss{O})_{h(n,k)r(n',k')} (\mathbfss{S})_{r(n',k')j(l,m,s)} \\
  \end{split}
\end{equation}
Where the subscript $R$ means that $\mathbfss{O}_{R}$ is the matrix representation with respect to
the real spherical harmonic basis. Eq.~\eqref{eq:basis-transformation-matrix-representation} in
matrix form reads $\mathbfss{O}_{R} = \mathbfss{S}^{\bm{\dagger}} \mathbfss{O} \mathbfss{S}$.

We now transform the complex system of equations~\eqref{eq:complex-system-of-equations} by
multiplying it with $\mathbfss{S}^{\bm{\dagger}}$ from the left and inserting the identity matrix
$\mathbfss{1} = \mathbfss{S}^{\bm{\dagger}}\mathbfss{S}$ where necessary. This yields
\begin{equation}
  \label{eq:real-system-of-equations}
  \begin{split}
    &\partial_{t}\mathbfss{S}^{\bm{\dagger}}\mathbfit{f}  +  \frac{V}{c^{2}} U_{a} \mathbfss{S}^{\bm{\dagger}}\mathbfss{A}^{a}\mathbfss{S}\partial_{t}\mathbfss{S}^{\bm{\dagger}}\mathbfit{f}
      + \left(U^{a}\mathbfss{1} + V \mathbfss{S}^{\bm{\dagger}}\mathbfss{A}^{a}\mathbfss{S}\right) \partial_{x_{a}}\mathbfss{S}^{\bm{\dagger}}\mathbfit{f}
    - \left(\gamma m \frac{\mathrm{d} U_{a}}{\mathrm{d} t} \mathbfss{S}^{\bm{\dagger}}\mathbfss{A}^{a}\mathbfss{S}
     + p \frac{\partial U_{b}}{\partial x^{a}} \mathbfss{S}^{\bm{\dagger}}\mathbfss{A}^{a}\mathbfss{S}\mathbfss{S}^{\bm{\dagger}}\mathbfss{A}^{b}\mathbfss{S}\right)\partial_{p} \mathbfss{S}^{\bm{\dagger}}\mathbfit{f} \\
    &\qquad\,\, {} + \left(\frac{\complexi}{V} \epsilon_{abc} \frac{\mathrm{d} U^{a}}{\mathrm{d} t} \mathbfss{S}^{\bm{\dagger}}\mathbfss{A}^{b}\mathbfss{S}\mathbfss{S}^{\bm{\dagger}}\bm{\Omega}^{c}\mathbfss{S}
    + \complexi \epsilon_{bcd} \frac{\partial U_{b}}{\partial x^{a}}\mathbfss{S}^{\bm{\dagger}}\mathbfss{A}^{a}\mathbfss{S}\mathbfss{S}^{\bm{\dagger}}\mathbfss{A}^{c}\mathbfss{S}\mathbfss{S}^{\bm{\dagger}}\bm{\Omega}^{d}\mathbfss{S}\right) \mathbfss{S}^{\bm{\dagger}}\mathbfit{f}
    - \complexi \omega_{a}\mathbfss{S}^{\bm{\dagger}}\bm{\Omega}^{a}\mathbfss{S}\mathbfss{S}^{\bm{\dagger}}\mathbfit{f}
      + \nu \mathbfss{S}^{\bm{\dagger}}\mathbfss{C}\mathbfss{S}\mathbfss{S}^{\bm{\dagger}}\mathbfit{f} \\
    &\qquad =  \left(\mathbfss{1} + \frac{V}{c^{2}} U_{a} \mathbfss{A}^{a}_{R} \right)\partial_{t}\mathbfit{f}_{R} 
      + \left(U^{a}\mathbfss{1} + V \mathbfss{A}^{a}_{R}\right) \partial_{x_{a}}\mathbfit{f}_{R}
    - \left(\gamma m \frac{\mathrm{d} U_{a}}{\mathrm{d} t} \mathbfss{A}^{a}_{R}
     + p \frac{\partial U_{b}}{\partial x^{a}} \mathbfss{A}^{a}_{R}\mathbfss{A}^{b}_{R}\right)\partial_{p} \mathbfit{f}_{R} \\
    &\qquad \,\, {}+ \left(\frac{1}{V} \epsilon_{abc} \frac{\mathrm{d} U^{a}}{\mathrm{d} t} \mathbfss{A}^{b}_{R}\bm{\Omega}^{c}_{R}
    + \epsilon_{bcd} \frac{\partial U_{b}}{\partial x^{a}}\mathbfss{A}^{a}_{R}\mathbfss{A}^{c}_{R}\bm{\Omega}^{d}\right) \mathbfit{f}_{R}
    - \omega_{a}\bm{\Omega}^{a}_{R}\mathbfit{f}_{R}
      + \nu \mathbfss{C}_{R}\mathbfit{f}_{R} = 0 \,.
  \end{split}
\end{equation}
Note in particular that
$\bm{\Omega}^{a}_{R} = \complexi \mathbfss{S}^{\bm{\dagger}}\bm{\Omega}^{a}\mathbfss{S}$ includes
the factor $\complexi$.

It is left to show that eq.~\eqref{eq:real-system-of-equations} is actually a real system of
equations, i.e. that all the appearing matrices and $\mathbfit{f}_{R}$ are real. We begin with the
latter and show that the components of $\mathbfit{f}_{R}$ are 
\begin{equation}
  \label{eq:real-expansion-coefficients}
  (\, \mathbfit{f}_{R})_{i(l,m,s)} = \sum_{l',m'} (\mathbfss{S}^{\bm{\dagger}})_{i(l,m,s)j(l',m')}(\, \mathbfit{f})^{\phantom{*}}_{j(l',m')}
  =  \frac{f^{m}_{l} + (-1)^{m}f^{-m}_{l}}{\sqrt{2(1 + \delta_{m0})}}\delta_{s0}
    + \frac{\complexi (f^{m}_{l} - (-1)^{m}f^{-m}_{l})}{\sqrt{2}}\delta_{s1} = f_{lms} \,.
\end{equation}
These are the expansion coefficients of the real spherical harmonic expansion, which we derived in
eq.~\eqref{eq:real-spherical-harmonic-expansion} and which, as we pointed out there, are real. To
prove that the matrices are real, we had to compute them explicitly. The explicit expressions for
all real representation matrices can be found in
Appendix~\ref{app:representation-matrices-real-spherical-harmonics}.

At the end of this section, we stress that the real representation matrices can as
well be computed with the help of rotation matrices, i.e.
\begin{alignat}{2}
  \label{eq:real-direction-op-matrices-and-rotations}
    \mathbfss{A}^{y}_{R}          & =
                           e^{-\frac{\upi}{2}\bm{\Omega}^{z}_{R}}\mathbfss{A}^{x}_{R}e^{\frac{\upi}{2}\bm{\Omega}^{z}_{R}}
                           \qquad & 
     \mathbfss{A}^{z}_{R}         & =
                           e^{-\frac{\upi}{2}\bm{\Omega}^{x}_{R}}\mathbfss{A}^{y}_{R}e^{\frac{\upi}{2}\bm{\Omega}^{x}_{R}} \quad \text{and} \\
  \label{eq:real-angular-momentum-op-matrices-and-rotations}
    \bm{\Omega}^{y}_{R}           & = e^{-\frac{\upi}{2}\bm{\Omega}^{z}_{R}}\bm{\Omega}^{x}_{R}e^{\frac{\upi}{2}\bm{\Omega}^{z}_{R}} & 
    \bm{\Omega}^{z}_{R}           & = e^{-\frac{\upi}{2}\bm{\Omega}^{x}_{R}}\bm{\Omega}^{y}_{R}e^{\frac{\upi}{2}\bm{\Omega}^{x}_{R}} \,.
\end{alignat}

\section{Eigenvectors and eigenvalues of the representation matrices}
\label{sec:eigenvectors-and-eigenvalues-representation matrices}

In the last section of this paper we investigate the eigenvectors and eigenvalues of the
representation matrices of the direction operators and the angular momentum operators. The former
play an important role in the implementation of numerical methods used to solve the system of
equations, which we have derived. Examples are the finite volume method or the discontinuous
Galerkin method. The statements about the eigenvalues of the representation matrices of the angular
momentum operators are included for the sake of completeness.

\subsection{General statements}
\label{sec:general-statements}

A first important observation is that the matrix representations of the $\operator{A}^{a}$ and
$\operator{L}^{a}$ are Hermitian matrices, because the eigenvalues of Hermitian matrices must be
real. Assume $\bm{\xi}$ is an eigenvector of an arbitrary Hermitian matrix $\mathbfss{A}$ with
eigenvalue $\lambda$ and with norm $ \xi \coloneqq \sqrt{\bm{\xi}^{*} \cdot \bm{\xi}} = 1$, then
\begin{equation}
  \label{eq:hermitian-matrices-real-eigenvalues}
  \lambda^{*} = \left(\bm{\xi}^{*} \cdot \mathbfss{A}\bm{\xi}\right)^{*}
  = (\bm{\xi})_{i}(\mathbfss{A})^{*}_{ij}(\bm{\xi})^{*}_{j}
  = (\bm{\xi})^{*}_{j} (\mathbfss{A}^{\transpose})^{*}_{ji}(\bm{\xi})_{i}
  = \bm{\xi}^{*} \cdot \mathbfss{A}^{\bm{\dagger}}\bm{\xi}
  = \bm{\xi}^{*} \cdot \mathbfss{A}\bm{\xi}
  =\lambda \,.
\end{equation}

We now prove the above observation for the direction operators $\operator{A}^{a}$. We prove it for
$\operator{A}^{x}$ and only note that proof for the other two operators works analogously. We use
the definition of the representation matrices given in eq.~\eqref{eq:advection-term} to get the
transpose conjugate representation matrix of $\operator{A}^{x}$, namely
\begin{equation}
  \label{eq:A-matrices-hermitian}
  (\mathbf{A}^{x\bm{\dagger}})_{i(l',m')j(l,m)} =
  {\scalarproduct{Y^{m}_{l}}{\operator{A}^{x}Y^{m'}_{l'}}}^{*}
  =  \left(\int_{S^{2}} {Y^{m}_{l}}^{*}\cos\theta Y^{m'}_{l'} \mathrm{d}\Omega\right)^{*}
  = \int_{S^{2}} {Y^{m'}_{l'}}^{*}\cos\theta Y^{m}_{l} \mathrm{d}\Omega
  = \scalarproduct{Y^{m'}_{l'}}{\operator{A}^{x}Y^{m}_{l}}
  =(\mathbf{A}^{x})_{i(l',m')j(l,m)} \,.
\end{equation}
Hence $\mathbfss{A}^{x}$ is a Hermitian matrix. The same is true for the representation matrices of
$\operator{A}^{y}$ and $\operator{A}^{z}$. Hence,
${\mathbfss{A}^{a}}^{\bm{\dagger}} = \mathbfss{A}^{a}$ for $a \in \{x, y, z\}$.

We proceed by proving that the representation matrices of the angular momentum operators are
Hermitian. Knowing their action on the spherical harmonics, see eq. \eqref{eq:action-L-operators},
we conclude that
\begin{equation}
  \label{eq:Omega-x-hermitian}
  (\bm{\Omega}^{x\bm{\dagger}})_{i(l',m')j(l,m)}
  = {\scalarproduct{Y^{m}_{l}}{\operator{L}^{x}Y^{m'}_{l'}}}^{*}
  = m' \delta_{mm'}
  = m \delta_{m'm}
  = \scalarproduct{Y^{m'}_{l'}}{\operator{L}^{x}Y^{m}_{l}}
  =(\bm{\Omega}^{x})_{i(l',m')j(l,m)} \,,  
\end{equation}
where we used that we get the same result when the roles of $m'$ and $m$ are interchanged. For
$\operator{L}^{y}$, we find
\begin{equation}
  \label{eq:Omega-y-hermitian}
  \begin{split}
    (\bm{\Omega}^{y\bm{\dagger}})_{i(l',m')j(l,m)}
    = {\scalarproduct{Y^{m}_{l}}{\operator{L}^{y}Y^{m'}_{l'}}}^{*}
    &= \frac{1}{2} \sqrt{(l' + m' + 1)(l' - m')} \delta_{ll'}\delta_{m(m'+1)}
      + \frac{1}{2} \sqrt{(l' + m') (l' - m' + 1)} \delta_{ll'} \delta_{m(m'-1)} \\
    &= \frac{1}{2} \sqrt{(l + m) (l - m + 1)} \delta_{l'l} \delta_{m'(m-1)}
      + \frac{1}{2} \sqrt{(l + m + 1)(l - m)} \delta_{l'l}\delta_{m'(m+1)} \\
    & = \scalarproduct{Y^{m'}_{l'}}{\operator{L}^{y}Y^{m}_{l}}
      = (\bm{\Omega}^{y})_{i(l',m')j(l,m)} \,.
  \end{split}
\end{equation}

The proof for $\operator{L}^{z}$ is analogous.

A second statement about the eigenvalues of the representation matrices follows from the fact that
the representation matrices are ``rotated'' versions of each other. This implies that their
eigenvalues are the same.

The proof requires the concept of similar matrices: We say the matrices $\mathbfss{X}$ and $\mathbfss{Y}$ are \textit{similar}, if they are related by
$\mathbfss{Y} = \mathbfss{T}^{-1} \mathbfss{X} \mathbfss{T}$. Similar matrices have the same
eigenvalues, because their characteristic polynomial is the same, i.e.
\begin{equation}
  \label{eq:eigenvalues-similar-matrices}
  p(\lambda) \coloneqq \det\left(\mathbfss{X} - \lambda \mathbfss{I}\right)
  = \det\left(\mathbfss{T} (\mathbfss{T}^{-1}\mathbfss{X} \mathbfss{T}  - \lambda \mathbfss{I}) \mathbfss{T}^{-1}\right)
  = \det(\mathbfss{T})\det(\mathbfss{Y} - \lambda\mathbfss{I})\det(\mathbfss{T}^{-1})
  = \det(\mathbfss{Y} - \lambda\mathbfss{I}) \,.
\end{equation}

Eq.~\eqref{eq:A-y-from-A-x} says that $\mathbfss{A}^{x}$ is similar to $\mathbfss{A}^{y}$ and
eq.~\eqref{eq:A-z-from-A-y} states that $\mathbfss{A}^{y}$ is similar to $\mathbfss{A}^{z}$. The
same is true for the representation matrices of the angular momentum operator, see
eq.~\eqref{eq:Omega-y-and-Omega-z-from-Omega-x}. Thus, all three matrices share the same
eigenvalues. We define the \textit{spectrum} to be the set of all eigenvalues of a matrix, namely
$\sigma(\mathbfss{X}) \coloneqq \{ \lambda \in \mathbb{C} \mid \exists\mathbfit{w} \neq 0 :
\mathbfss{X}\mathbfit{w} = \lambda\mathbfit{w} \}$. The fact that all three matrices have the same
eigenvalues can now be expressed as
\begin{equation}
  \label{eq:definition-spectrum}
  \sigma(\mathbfss{A}^{x}) = \sigma(\mathbfss{A}^{y}) = \sigma(\mathbfss{A}^{z}) \,.
\end{equation}

A third statement concerns the eigenvectors of the representation matrices: The eigenvectors are
rotated versions of each other, i.e.
\begin{equation}
  \label{eq:eigenvectors-A-x-A-y-A-z}
  \mathbfss{A}^{x}\mathbfss{W}^{x} = \mathbfss{W}^{x}\bm{\Lambda} \iff
  e^{-i\frac{\upi}{2}\bm{\Omega}^{z}} \mathbfss{A}^{x} e^{i\frac{\upi}{2}\bm{\Omega}^{z}} e^{-i\frac{\upi}{2}\bm{\Omega}^{z}}\mathbfss{W}^{x} = e^{-i\frac{\upi}{2}\bm{\Omega}^{z}}\mathbfss{W}^{x}\bm{\Lambda}
  \iff
  \mathbfss{A}^{y } e^{-i\frac{\upi}{2}\bm{\Omega}^{z}}\mathbfss{W}^{x} = e^{-i\frac{\upi}{2}\bm{\Omega}^{z}}\mathbfss{W}^{x}\bm{\Lambda}
  \Longleftrightarrow \mathbfss{A}^{y}\mathbfss{W}^{y} = \mathbfss{W}^{y}\bm{\Lambda} \,.
\end{equation}
Where $\bm{\Lambda}$ is a diagonal matrix with the eigenvalues of $\mathbfss{A}^{x}$ on its diagonal
and $\mathbfss{W}$ are matrices whose columns are the corresponding eigenvectors. An analogue
computation can be performed for $\mathbfss{A}^{z}$.

\subsection{Eigenvalues of the representation matrices of the direction operators}
\label{sec:eigenvalues-direction-operator-matrices}

In the previous section we showed that if we know the eigenvalues of one of the representation
matrices, we know them for all. Since $\mathbfss{A}^{x}$ is simpler, i.e. it has less non-zero
elements, we focus on determining its eigenvalues and, as we will see, its eigenvalues are the roots
of the associated Legendre polynomials.

To see a general pattern emerge, we begin with computing the eigenvalues of $\mathbfss{A}^{x}$ for
$l_{\text{max}} = 1$ and $l_{\text{max}} = 2$. Using the following ordering of the spherical
harmonics $\mathcal{S}^{1}=\text{span}\{Y^{0}_{0},Y^{0}_{1},Y^{1}_{1},Y^{-1}_{1}\}$ for
$l_{\text{max}} = 1$ and
$\mathcal{S}^{2}=\text{span}\{Y^{0}_{0},Y^{0}_{1},Y^{0}_{2}, Y^{1}_{1}, Y^{1}_{2}, Y^{-1}_{1},
Y^{-1}_{2}, Y^{2}_{2},Y^{-2}_{2}\}$ for $l_{\text{max}} = 2$, the matrix $\mathbfss{A}^{x}$ is, as
we will see now, tridiagonal. We use eq.~\eqref{eq:Ax} to compute its matrix elements and introduce
the shorthand $c^{m}_{l} = [(l + m) (l - m)/(2l + 1) (2l - 1)]^{1/2}$ for the coefficients in the
referred equation. This yields
\begin{equation}
  \label{eq:A-x-lequal1-lequal2}
  \begin{split}
    \mathbfss{A}^{x}           & =
    \begin{pmatrix}
      0                        & c^{0}_{1}              & 0                       & 0                                                                     \\
      c^{0}_{1}                & 0                      & 0                       & 0                                                                     \\
      0                        & 0                      & 0                       & 0                                                                     \\
      0                        & 0                      & 0                       & 0 
    \end{pmatrix} \coloneqq
    \begin{pmatrix}
      \mathbfss{A}^{x}_{1,0}   &                        &                                                                                                 \\
                               & \mathbfss{A}^{x}_{1,1} &                                                                                                 \\
                               &                        & \mathbfss{A}^{x}_{1,-1}
    \end{pmatrix}
      \;\text{ for $l_{\rm max} = 1$ and}\;                                                                                                                                        \\
    \mathbfss{A}^{x}           & =
    \begin{pmatrix}
      0                        & c^{0}_{1}              & 0                       &                           &           &            &            &   & \\
      c^{0}_{1}                & 0                      & c^{0}_{2}               &                           &           &            &            &   & \\
      0                        & c^{0}_{2}              & 0                       &                           &           &            &            &   & \\
                               &                        &                         & 0                         & c^{1}_{2} &            &            &   & \\
                               &                        &                         & c^{1}_{2}                 & 0         &            &            &   & \\
                               &                        &                         &                           &           & 0          & c^{-1}_{2} &   & \\
                               &                        &                         &                           &           & c^{-1}_{2} & 0          &   & \\
                               &                        &                         &                           &           &            &            & 0 & \\
                               &                        &                         &                           &           &            &            &   & 0 
    \end{pmatrix}
    \coloneqq
    \begin{pmatrix}
      \mathbfss{A}^{x}_{2,0}   &                        &                         &                           &           &                               \\
                               & \mathbfss{A}^{x}_{2,1} &                         &                           &                                           \\
                               &                        & \mathbfss{A}^{x}_{2,-1} &                           &           &                               \\
                               &                        &                         & \mathbfss{A}^{x}_{2,2}    &           &                               \\
                               &                        &                         &                           & \mathbfss{A}^{x}_{2,-2}
    \end{pmatrix} \;\text{for $l_{\rm max} = 2$}\,.
  \end{split}
\end{equation}
Where we introduced the tridiagonal block matrices
\begin{equation}
  \label{eq:block-matrices}
  \mathbfss{A}^{x}_{l,m} =
  \begin{pmatrix}
    0               & c^{m}_{|m| + 1} &        &                        &                        \\
    c^{m}_{|m| + 1} & 0               & c^{m}_{|m| + 2}                                          \\
                    & c^{m}_{|m| + 2} & \ddots & \ddots                                          \\
                    &                 & \ddots &  \ddots                      & c^{m}_{l} \\
                    &                 &        & c^{m}_{l} & 0
  \end{pmatrix}
  \in \mathbb{R}^{(l-m + 1) \times(l - m + 1) } \text{ with } |m| \leq l \text{ and } \mathbfss{A}^{x}_{l,\pm l} \coloneqq 0 .
\end{equation}
We emphasise that the size of the block matrices $\mathbfss{A}^{x}_{l, m}$ varies with $l$ and that
the pattern in eq.~\eqref{eq:A-x-lequal1-lequal2} is the same for arbitrary $l_{\text{max}}$.

The block diagonal structure of $\mathbfss{A}^{x}$ implies that its characteristic polynomial
factors into the characteristic polynomials of its blocks, i.e.
\begin{equation}
  \label{eq:characterstic-polynomial-A-x}
  p(\lambda) = \det(\mathbfss{A}^{x} - \lambda\mathbfss{1}) = \prod^{l_{\text{max}}}_{m =-l_{\text{max}}}\det(\mathbfss{A}^{x}_{l_{\text{max}},m} - \lambda \mathbfss{1})
  = \det(\mathbfss{A}^{x}_{l_{\text{max}},0} - \lambda \mathbfss{1}) \prod^{l_{\text{max}}}_{m = 1} \left[\det(\mathbfss{A}^{x}_{l_{\text{max}},m} - \lambda \mathbfss{1})\right]^{2}\,.
\end{equation}
Where we used that $\mathbfss{A}_{l_{\text{max}}, m} = \mathbfss{A}_{l_{\text{max}},-m}$, because
$c^{m}_{l} = c^{-m}_{l}$.

A direct consequence of eq.~\eqref{eq:characterstic-polynomial-A-x} is that the eigenvalues of
$\mathbfss{A}^{x}$ (i.e. the roots of its characteristic polynomial) are the eigenvalues of the
block matrices $A^{x}_{l_{\text{max}, m}}$. We now prove that their eigenvalues are the roots of the
$m$th derivative of the Legendre polynomial $P_{l+1}$.

Before we begin the proof we introduce 
\begin{equation}
  \label{eq:polynomial-inside-the-associated}
  p^{m}_{l}(x) \coloneqq \frac{\mathrm{d}^{m}}{\mathrm{d}x^{m}} P_{l}(x) \,,
\end{equation}
to denote the $m$th derivative of the Legendre polynomial. Furthermore, we point out that
$p^{m}_{l}$ is part of the definition of the associated Legendre polynomials, i.e.
$P^{m}_{l}(x) = (-1)^{m} (1 - x^{2})^{m/2} p^{m}_{l}(x)$. Thus, if the eigenvalues of
$\mathbfss{A}^{x}_{l,m}$ are the roots of the $m$th derivative of the Legendre polynomial $P_{l+1}$,
they are as well the roots of the associated Legendre polynomial $P^{m}_{l+1}$ modulo $\pm 1$.

In a first step, we compute $\det(\mathbfss{A}^{x}_{l,m} - \lambda \mathbfss{1})$ developing it
after the last row, i.e.
\begin{equation}
  \label{eq:det-A-l-m-x}
 \pi^{m}_{l+1} (\lambda) \coloneqq \det(\mathbfss{A}^{x}_{l,m} - \lambda \mathbfss{1})
 = -\lambda \det(\mathbfss{A}^{x}_{l-1,m} - \lambda \mathbfss{1}) - ({c^{m}_{l}})^{2} \det(\mathbfss{A}^{x}_{l-2,m} - \lambda \mathbfss{I})
 = -\lambda \pi^{m}_{l}(\lambda) - ({c^{m}_{l}})^{2} \pi^{m}_{l-1}(\lambda) \,.
\end{equation}

Eq.~\eqref{eq:det-A-l-m-x} can be read as a recurrence relation for the polynomials $\pi^{m}_{l}$.
Fixing $m$ to some integer value and setting $\pi^{m}_{|m|}(\lambda) \coloneqq 1$ the characteristic
polynomials for arbitrary $l,m$ can be computed recursively.

In a second step, we compare the recurrence relation~\eqref{eq:det-A-l-m-x} with the recurrence
relation for the associated Legendre polynomials in eq.~\eqref{eq:recurrence_relation_xP}, which
implies a recurrence relation for the $m$th derivative of the Legendre polynomials, i.e.
\begin{equation}
  \label{eq:rearranged-recurrence-relation-xPlm}
     P^{m}_{l+1}(x)  =
      x \frac{2l + 1}{l - m + 1} P^{m}_{l}(x) - \frac{(l + m)}{(l - m + 1)} P^{m}_{l-1}(x)
     \implies      p^{m}_{l+1}(x)  =
      x \frac{2l + 1}{l - m + 1} p^{m}_{l}(x) - \frac{(l + m)}{(l - m + 1)} p^{m}_{l-1}(x) \,.
\end{equation}
We see that the recurrence relations in eq.~\eqref{eq:det-A-l-m-x} and
eq.~\eqref{eq:rearranged-recurrence-relation-xPlm} are similar. They differ in the factors in front
of the polynomials on the right-hand side. This suggests that $p^{m}_{l}(x) \propto \pi^{m}_{l}(x)$.
If the polynomials are proportional to each other, they have the same roots; multiplying a function
with a constant does not change its roots. The factor of proportionality can be derived by noting
that the leading order term of $\pi^{m}_{l}(x)$ is $(-1)^{l-m}x^{l-m} $, using an explicit formula
for the associated Legendre polynomials and dividing its leading order term by the factor which
accompanies it. This results in
\begin{equation}
  \label{eq:P-l-m-proportional-characteristic-polynomial}
  p^{m}_{l}(x) = (-1)^{l} \frac{(2l - 1)!!}{(l-m)!} \pi^{m}_{l}(x) \, .
\end{equation}

We summarise and conclude that the eigenvalues of the representation matrices of the operators
$\operator{A}^{a}$ for a fixed $l_{\text{max}}$ are the roots of the polynomials
$\pi^{m}_{l_{\text{max}}+1}$ with $ 0 \leq m \leq l_{\text{max}}$, which are, modulo a constant of
proportionality, the $m$th derivatives of the Legendre Polynomial $P_{l+1}$. Hence, the eigenvalues
of $\mathbfss{A}^{a}$ are the roots of the associated Legendre Polynomials
$P^{0}_{l_{\text{max}} + 1}, \dots, P^{l_{\text{max}}}_{l_{\text{max}} + 1}$ modulo $\pm 1$. We note
that the roots of the polynomials with $m \neq 0$ have algebraic multiplicity two, see eq.
\eqref{eq:characterstic-polynomial-A-x}.

The above conclusions have implications for the possible set of eigenvalues: First, the Legendre
polynomials are orthogonal polynomials and the roots of orthogonal polynomials are real, simple and
located in the interior of the interval of orthogonality, which is $[-1, 1]$ in the case of the
Legendre polynomials, see~\citet[Section 22.16]{Stegun_HandbookOfFunctions}. Secondly, the roots of
$P^{m}_{l}$ with $m \neq 0$ are the roots of the derivatives of the Legendre polynomials, i.e. the
roots of $p^{m}_{l}$. Rolle's theorem states that the roots of the derivative of a continuous
and differentiable function, must lie between the roots of this function. This implies that the
roots of the polynomials $p^{m}_{l}$ are ``moving'' towards zero with increasing $m$, hence that all
eigenvalues of $\mathbfss{A}^{x}$ are contained in the interval $[-1, 1]$ and that the largest
eigenvalue of $\mathbfss{A}^{x}$ is the largest root of the Legendre polynomial
$P^{0}_{l_{\text{max}} + 1} = P_{l_{\text{max}} + 1}$.

We can get a very good estimate for the largest eigenvalue using a formula to compute estimates of
the roots of the Legendre polynomial of degree $l$, which can be found
in~\citet[eq.13]{Tricomi1950}. The formula is
\begin{equation}
  \label{eq:estimate-roots-P-l}
  x_{r} = \left(1 - \frac{1}{8l^{2}} + \frac{1}{8l^{3}}\right)
  \cos\left(\frac{4r - 1}{4l + 2} \upi\right) + \mathcal{O}(l^{-4})\quad \text{ with } r \in
  \begin{cases}
    \{1, \dots, l/2\} &\text{if $l$ even}\\
    \{1, \dots, (l - 1)/2\} &\text{if $l$ odd}
  \end{cases} \,.
\end{equation}
Note that the Legendre polynomials are even (odd) for even (odd) $l$. The estimate
for the largest eigenvalue of $\mathbfss{A}^{x}$ is $\lambda_{\text{max}} \approx x_{1}$.

\subsection{Eigenvalues and eigenvectors of the sum of representation matrices}
\label{sec:eigenvectors-eigenvalues-sum-of-matrices}

In the last part of this section we show how to compute the eigenvalues and eigenvectors of
\begin{equation}
  \label{eq:sum-A-matrices}
  \eta_{x}\mathbfss{A}^{x} + \eta_{y}\mathbfss{A}^{y} +
  \eta_{z}\mathbfss{A}^{z} \quad \text{where } \bm{\eta} = (\eta_{x}, \eta_{y}, \eta_{z})^{T} \in \mathbb{R}^{3}\,
\end{equation}
using the eigenvalues and eigenvectors of $\mathbfss{A}^{x}$.

In a first step we note that the sum of the representation matrices of the direction operators can
be traced back to the scalar product
$\bm{\eta} \cdot \mathbfit{e}_{p} = \eta_{x} \operator{A}^{x} + \eta_{y} \operator{A}^{y} + \eta_{z}
\operator{A}^{z}$, see eq.~\eqref{eq:nabla-p-momentum-part-A}. Secondly, we define the unit vector
$\mathbfit{n} \coloneqq \bm{\eta}/\eta$. Thirdly, we rotate the coordinate system such that the
$x$-axis of the rotated coordinate system is parallel to $\mathbfit{n}$. We denote the corresponding
rotation matrix with $\mathbfss{R}_{\mathbfit{n}}$. This changes the scalar product
$\bm{\eta} \cdot \mathbfit{e}_{p}$ to
\begin{equation}
  \label{eq:change-of-scalar-product-under-rotation}
  \eta \mathbfit{n} \cdot \left(\mathbfss{R}^{\transpose}_{\mathbfit{n}} \mathbfss{R}_{\mathbfit{n}}\mathbfit{e}_{p}\right)
  = \eta \left(\mathbfss{R}_{\mathbfit{n}}\mathbfit{n}\right) \cdot \left( \mathbfss{R}_{\mathbfit{n}}\mathbfit{e}_{p}\right)
  = \eta \tilde{\bm{n}} \cdot \tilde{\mathbfit{e}}_{p}
  = \eta \mathbfit{e}_{x} \cdot \tilde{\mathbfit{e}}_{p}
  = \eta \cos{\tilde{\theta}}
  = \eta \tilde{A}_{x} \,,
\end{equation}
where we used that the coordinates $\tilde{\mathbfit{n}}$ of $\mathbfit{n}$ in the rotated
coordinate system are $\mathbfit{e}_{x}$. Eq.~\eqref{eq:change-of-scalar-product-under-rotation}
shows that the sum of the direction operators given by the scalar product
$\bm{\eta} \cdot \mathbfit{e}_{p}$ reduces to $\eta \tilde{A}_{x}$ in the rotated coordinate system
and so does its matrix representation~\eqref{eq:sum-A-matrices} with respect to the spherical
harmonics defined in it.

As explained in Section~\ref{sec:rotations-spherical-harmonics} transforming representation matrices
into a rotated coordinate system requires to know how to rotate spherical harmonics. In
eq.~\eqref{eq:wigner-d-functions} we presented the necessary rotation matrix
$\mathbfss{U}(\alpha \mathbfit{r}) = e^{-\complexi \alpha r_{a} \bm{\Omega}^{a}}$, where the unit
vector $\mathbfit{r}$ is the axis of rotation and $\alpha$ is the angle by which the spherical
harmonics are rotated.

Because of eq.~\eqref{eq:change-representation-matrix-rotation} and since we know that the sum of
matrices~\eqref{eq:sum-A-matrices} must equal $\eta \mathbfss{A}^{x}$ in the rotated frame,
\begin{equation}
  \label{eq:eta-A-x-sum-of-matrices}
  \eta \mathbfss{A}^{x} = \mathbfss{U}(\alpha \mathbfit{r})^{\bm{\dagger}}(  \eta_{x}\mathbfss{A}^{x} + \eta_{y}\mathbfss{A}^{y} +
  \eta_{z}\mathbfss{A}^{z}) \mathbfss{U}(\alpha \mathbfit{r}) \,.
\end{equation}
The axis of rotation can be computed with
$\mathbfit{r} = \mathbfit{e}_{x} \times \mathbfit{n} /\| \mathbfit{e}_{x} \times \mathbfit{n}\|$ and
the angle $\alpha = \mathbfit{e}_{x} \cdot \mathbfit{n}$.

We can now investigate the eigenvalues of the sum of the representation
matrices~\eqref{eq:sum-A-matrices}. They are determined by
\begin{equation}
  \label{eq:eigenvalues-sum-of-representation-matrices}
  \begin{split}
    \det\left(\eta_{x}\mathbfss{A}^{x} + \eta_{y}\mathbfss{A}^{y} +
    \eta_{z}\mathbfss{A}^{z} - \lambda^{s}\mathbfss{1}\right)
    &=   \det\left(\mathbfss{U}(\alpha \mathbfit{r})\left(\mathbfss{U}^{\bm{\dagger}}(\alpha  \mathbfit{r})\left(\eta_{x}\mathbfss{A}^{x} + \eta_{y}\mathbfss{A}^{y} +
      \eta_{z}\mathbfss{A}^{z}\right)\mathbfss{U}(\alpha \mathbfit{r}) - \lambda^{s}\mathbfss{1}\right) \mathbfss{U}^{\bm{\dagger}}(\alpha \mathbfit{r})\right) 
    =    \det\left(\eta \mathbfss{A}^{x} - \lambda^{s}\mathbfss{1}\right) = 0 \, , 
  \end{split}
\end{equation}
whence we conclude that the eigenvalues of the sum of the representation
matrices~\eqref{eq:sum-A-matrices} are the eigenvalues of $\mathbfss{A}^{x}$ times $\eta$, i.e. its
$i$th eigenvalue is
\begin{equation}
  \label{eq:ith-eigenvalue-of-matrix-sum}
  \lambda^{s}_{i} = \lambda_{i}(\eta^{2}_{x} + \eta^{2}_{y} + \eta^{2}_{z})^{1/2}
  \quad \text{where $\lambda_{i}$ is the $i$th eigenvalue of $\mathbfss{A}^{x}$} \,.
\end{equation}
We highlight that this relation between the eigenvalues of $\mathbfss{A}_{x}$ and the eigenvalues of
the sum of the representation matrices is independent of $l_{\text{max}}$ and holds for each
eigenvalue $\lambda^{s}_{i}$.

\begin{figure}\centering
  \includegraphics{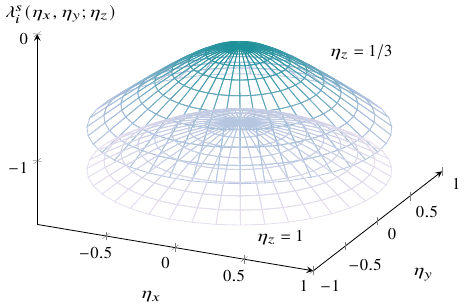}
  \caption{Plot of the eigenvalue $\lambda^{s}_{i}$ of
    $\eta_{x}\mathbfss{A}^{x} + \eta_{y}\mathbfss{A}^{y} + \eta_{z}\mathbfss{A}^{z}$ for varying
    $\eta_{x}$ and $\eta_{y}$ and fixed $\eta_{z}$. The plotted $\lambda^{s}_{i}$ corresponds to the
    eigenvalue $\lambda_{i} = -0.9062$ of $\mathbfss{A}^{x}$ with $l_{\text{max}} = 4$ . The
    depicted circular hyperboloid is described by
    $ (\eta^{2}_{x} + \eta^{2}_{y})/\eta^{2}_{z} - (\lambda^{s}_{i}/\lambda_{i}\eta_{z})^{2} = -1$.
  }
  \label{fig:eigenvalues-matrix-sum}
\end{figure}

The geometrical interpretation of eq.~\eqref{eq:ith-eigenvalue-of-matrix-sum} is that for a given
value $\eta_{z}$ the eigenvalues $\lambda^{s}_{i}$ lie on one of the sheets of a circular and
two-sheeted hyperboloid which is parameterised by $\eta_{x}$ and $\eta_{y}$. To illustrate this we
plotted one eigenvalue of the sum of the representation matrices~\eqref{eq:sum-A-matrices} for two
values of $\eta_{z}$, see Fig.~\ref{fig:eigenvalues-matrix-sum}.

The eigenvectors of the sum of the representation matrices~\eqref{eq:sum-A-matrices} are the rotated
eigenvectors of $\mathbfss{A}^{x}$, because
\begin{equation}
  \label{eq:eigenvectors-matrix-sum}
  \eta \mathbfss{A}^{x}\mathbfss{W}^{x} = \eta \mathbfss{W}^{x}\bm{\Lambda}
  \iff
  \eta  \mathbfss{U}(\alpha \mathbfit{r})\mathbfss{A}^{x}\mathbfss{U}^{\bm{\dagger}}(\alpha \mathbfit{r})\mathbfss{U}(\alpha \mathbfit{r}) \mathbfss{W}^{x} = \eta \mathbfss{U}(\alpha \mathbfit{r})\mathbfss{W}^{x}\bm{\Lambda}
  \iff
  \left(\eta_{x}\mathbfss{A}^{x} + \eta_{y}\mathbfss{A}^{y} +  \eta_{z}\mathbfss{A}^{z}\right)\mathbfss{U}(\alpha \mathbfit{r}) \mathbfss{W}^{x}
  = \mathbfss{U}(\alpha \mathbfit{r})\mathbfss{W}^{x}\bm{\Lambda}^{s} \,.
\end{equation}

We conclude this section with referring the reader to \citet{Garret2016}, who also investigated the
eigenstructure of the $\mathbfss{A}^{a}$ matrices in the context of radiation transport and found
similar results. Our work differs from theirs mainly in identifying the unitary matrices
$\mathbfss{U}$ as rotation matrices. Furthermore we would like to remark, that the eigenvalues of
the real matrices $\mathbfss{A}^{a}_{R}$ are the same as the eigenvalues of the $\mathbfss{A}^{a}$
matrices, because $\mathbfss{S}^{\bm{\dagger}} \mathbfss{A}^{a} \mathbfss{S} = \mathbfss{A}^{a}_{R}$ is a
similarity transformation.

\section{Conclusions}
\label{sec:conclusions}

The transport of particles or photons is very often modelled using variants of the Boltzmann
equation. In many cases the scattering (or collision) operator has a particular simple form in the
centre-of-momentum frame. This motivates to formulate the transport equation in a mixed-coordinate
frame, i.e. a frame in which time and the configuration space variable $\mathbfit{x}$ are measured
in an inertial frame whereas $\mathbfit{p}$ is given with respect to a non-inertial frame moving
with the centre-of-momentum of the scattering centres. To render the equation amenable to be solved
numerically, the single particle phase-space density is expanded in terms of known functions, for
example the Legendre polynomials, to reduce the number of independent variables. This expansion
leads to a system of partial differential equations evolving the expansion coefficients.

In this paper, a new method to derive this system of equations for the case of a spherical harmonic
expansion of the single particle phase-space density has been presented. The method is based on operators
acting on the space of spherical harmonics and their matrix representations. It was shown that the
system is readily obtained by replacing the introduced and identified operators appearing in the VFP
equation with their matrix representations. The final result is a compact matrix equation for the
(real) expansion coefficients, which has the same structure and physical interpretation as the
original VFP equation, i.e.
\begin{equation}
\label{eq:real-system-of-equations_repeat}
\begin{split}
&\qquad  \left(\mathbfss{1} + \frac{V}{c^{2}} U_{a} \mathbfss{A}^{a}_{R} \right)\partial_{t}\mathbfit{f}_{R} 
+ \left(U^{a}\mathbfss{1} + V \mathbfss{A}^{a}_{R}\right) \partial_{x_{a}}\mathbfit{f}_{R}
- \left(\gamma m \frac{\mathrm{d} U_{a}}{\mathrm{d} t} \mathbfss{A}^{a}_{R}
+ p \frac{\partial U_{b}}{\partial x^{a}} \mathbfss{A}^{a}_{R}\mathbfss{A}^{b}_{R}\right)\partial_{p} \mathbfit{f}_{R} \\
&\qquad \,\,\,\,\, {}+ \left(\frac{1}{V} \epsilon_{abc} \frac{\mathrm{d} U^{a}}{\mathrm{d} t} \mathbfss{A}^{b}_{R}\bm{\Omega}^{c}_{R}
+ \epsilon_{bcd} \frac{\partial U_{b}}{\partial x^{a}}\mathbfss{A}^{a}_{R}\mathbfss{A}^{c}_{R}\bm{\Omega}^{d}\right) \mathbfit{f}_{R}
- \omega_{a}\bm{\Omega}^{a}_{R}\mathbfit{f}_{R}
+ \nu \mathbfss{C}_{R}\mathbfit{f}_{R} = 0 \,.
\end{split}
\end{equation}

We simplified the computation of the matrix representations considerable with the help of rotations
of the coordinate system in momentum space, see in particular eqs.~\eqref{eq:A-z-from-A-y}, \eqref{eq:A-y-from-A-x} and \eqref{eq:Omega-y-and-Omega-z-from-Omega-x}. We showed that the system of equations can be
constructed from four representation matrices of whom two are diagonal and well-known. 

The fact that the spherical harmonics (and the expansion coefficients) are complex whereas the
single particle phase-space density is real, implies that the constructed system of equations
contains linear dependent equations. We removed the redundant equations by multiplying the system
with a basis transformation matrix, which turns the complex spherical harmonics into the real
spherical harmonics, see eq.~\eqref{eq:real-system-of-equations}.

Eventually, we prepared the application of numerical methods, like the discontinuous Galerkin method
or the finite volume method, through an investigation of the eigenvalues and eigenvectors of the
representation matrices. A key insight consists in the fact that all representation matrices of the
direction operators have the same eigenvalues. This holds true for the representations matrices of
the angular momentum operators as well. Moreover, all eigenvalues are real. This was shown in Section~\eqref{sec:general-statements}. We emphasise, that we
have proven that the eigenvalues of the representation matrices of the direction operators are the
roots of the associated Legendre polynomials, see Section~\ref{sec:eigenvalues-direction-operator-matrices}. We were also able to directly compute the eigenvalues and eigenvectors of the sum of the representation matrices of the direction operators, see eq.~\eqref{eq:ith-eigenvalue-of-matrix-sum} and eq.~\eqref{eq:eigenvectors-matrix-sum}. These results have all been exploited in the implementation of the Sapphire++ code, a discontinuous Galerkin scheme that solves equation (\ref{eq:real-system-of-equations_repeat}). The details are described in a companion paper.

We conclude with two remarks: First, the presented method can be applied whenever a spherical
harmonic expansion of the single particle phase-space density function is used. In particular, it is
\emph{not} necessary to formulate the transport equation in a mixed-coordinate frame. The only
prerequisite is the ability to compute the representation matrix of the collision operator
$\operator{C}$. Secondly, the presented operator method could be adapted to other
expansions of the single particle phase-space density function as well.

\section*{Acknowledgements}

We thank the referee Prof. Gary Webb for providing comments on the manuscript. We are very grateful to Prof. John Kirk for his valuable contributions to our discussions. Moreover, we thank Florian Schulze for profoundly testing and discussing the proposed method.

\section*{Data Availability}

Data sharing not applicable to this article as no datasets were generated or analysed during the current study.

\section*{Supplementary material}

A free and open-source python script, which may serve as a reference implementation of the representation matrices, is available as supplementary material and at \url{https://github.com/nils-schween/representation-matrices}.



\bibliographystyle{mnras}
\bibliography{references} 




\appendix
\section{Spherical harmonics}
\label{app:spherical-harmonics}

We would like to explicitly define the spherical harmonics, namely
\begin{equation}
  \label{eq:spherical-harmonics}
  Y^{m}_{l}(\theta, \varphi) = N^{m}_{l} P^{m}_{l}(\cos\theta)e^{\complexi m \varphi}
  \quad \text{with }   N^{m}_{l} = \sqrt{\frac{2l + 1}{4\pi} \frac{(l-m)!}{(l+m)!}} \,.  
\end{equation}

The functions $P^{m}_{l}$ are the associated Legendre Polynomials and we decided to use the
following definition:
\begin{equation}
  \label{eq:associated-legendre-polynomials}
  P^{m}_{l}(\cos\theta) =
  (-1)^{m}\sin^{m}\theta \frac{\mathrm{d}^{m}}{\mathrm{d}(\cos\theta)^{m}}P_{l}(\cos\theta) \,.
\end{equation}

Note that the Condon--Shortley phase $(-1)^{m}$ is included in the definition of the associated
Legendre polynomials and \emph{not} in the definition of the spherical harmonics. The $P_{l}$ are
the Legendre polynomials, which for example are defined to be the set of orthogonal polynomials on
the interval $[-1,1]$ with the weight function $w(x) = 1$, and can be constructed by an application
of the Gram-Schmidt process to the monomials.

\section{Real representation matrices}
\label{app:representation-matrices-real-spherical-harmonics}

At the end of Section~\eqref{sec:turning-complex-system-into-real-system}, we referred the reader to
the Appendix for a proof that the representation matrices corresponding to the real spherical
harmonics are real.

We prove this statement by explicitly computing all the appearing matrices, i.e. we evaluate the
matrix-matrix products in $\mathbfss{S}^{\bm{\dagger}}\mathbfss{O} \mathbfss{S}$. Where
$\mathbfss{O}$ is a representation matrix with respect to the complex spherical harmonics of an
arbitrary operator.

The matrix elements of the inverse basis transformation are
\begin{equation}
  \label{eq:inverse-basis-transformation-matrix}
  \begin{split}
    (\mathbfss{S}^{\bm{\dagger}})_{i(l',m',s')j(l,m)}
    &= \delta_{ll'} \left(\frac{\delta_{mm'} + (-1)^{m'} \delta_{-m'm}}{\sqrt{2(1 + \delta_{m'0})}}\delta_{s'0} +  \frac{\complexi (\delta_{mm'}  -  (-1)^{m'}\delta_{-m'm})}{\sqrt{2}}  \delta_{s'1}\right) \,.
  \end{split}
\end{equation}

We now demonstrate how to compute the matrix representation of an arbitrary operator $\operator{O}$
corresponding to the real spherical harmonics. We assume that we know its matrix representation with
the respect to the complex spherical harmonics. Evaluating the necessary matrix-matrix products
yields
\begin{equation}
  \label{eq:real-representation-matrix-arbitrary-operator}
  \begin{split}
    &(\mathbfss{O}_{R})_{i(l',m',s')j(l,m,s)} \\
    &\qquad = \sum_{n,k} \sum_{n',k'} (\mathbfss{S}^{\bm{\dagger}})_{i(l',m',s')h(n,k)} \mathbfss{O}_{h(n,k)r(n',k')} (\mathbfss{S})_{r(n',k')j(l,m,s)} \\
                                                 &\qquad =  \sum_{n,k} \sum_{n',k'} \delta_{l'n} \delta_{ln'}\left(
                                                   \frac{\delta_{m'k} + (-1)^{m'}\delta_{-m'k}}{\sqrt{2(1 + \delta_{m'0})}} \delta_{s'0}
                                                   + \frac{\complexi(\delta_{m'k} - (-1)^{m'}\delta_{-m'k})}{\sqrt{2}} \delta_{s'1} \right) \mathbfss{O}_{h(n,k)r(n',k')}\\
    &\qquad\phantom{=} {} \times \left(\frac{\delta_{mk'} + (-1)^{m} \delta_{-mk'}}{\sqrt{2(1 + \delta_{m0})}}\delta_{s0} -  \frac{\complexi (\delta_{mk'}  -  (-1)^{m}\delta_{-mk'})}{\sqrt{2}})  \delta_{s1}\right) \\
                                                    &\qquad= \frac{1}{2}
                                                       \frac{\delta_{s'0}\delta_{s0}}{\sqrt{(1 + \delta_{m'0})(1 + \delta_{m0})}}  \left(\mathbfss{O}_{i(l',m')j(l,m)} + (-1)^{m'}\mathbfss{O}_{i(l',-m')j(l,m)}
                                                       + (-1)^{m} \mathbfss{O}_{i(l',m')j(l,-m)} + (-1)^{m'+ m} \mathbfss{O}_{i(l',-m')j(l,-m)} 
                                                      \right)\\
&\qquad\phantom{=} {}-                                                   \frac{\complexi}{2} \frac{\delta_{s'0}\delta_{s1}}{\sqrt{(1 + \delta_{m'0})}}  \left(\mathbfss{O}_{i(l',m')j(l,m)} + (-1)^{m'}\mathbfss{O}_{i(l',-m')j(l,m)}
                                                       - (-1)^{m} \mathbfss{O}_{i(l',m')j(l,-m)} - (-1)^{m' + m} \mathbfss{O}_{i(l',-m')j(l,-m)} 
  \right)\\
&\qquad\phantom{=} {} +              \frac{\complexi}{2} \frac{\delta_{s'1}\delta_{s0}}{\sqrt{(1 + \delta_{m0})}}  \left(\mathbfss{O}_{i(l',m')j(l,m)} - (-1)^{m'}\mathbfss{O}_{i(l',-m')j(l,m)} 
                                                       + (-1)^{m} \mathbfss{O}_{i(l',m')j(l,-m)} - (-1)^{m' + m} \mathbfss{O}_{i(l',-m')j(l,-m)} 
  \right)\\
&\qquad\phantom{=} {} +             \frac{1}{2} \delta_{s'1}\delta_{s1}\left(\mathbfss{O}_{i(l',m')j(l,m)} - (-1)^{m'}\mathbfss{O}_{i(l',-m')j(l,m)}
                                                       - (-1)^{m} \mathbfss{O}_{i(l',m')j(l,-m)} + (-1)^{m' + m} \mathbfss{O}_{i(l',-m')j(l,-m)} 
                                                       \right) \,.\\
  \end{split}
\end{equation}

Having derived an expression for the real matrix representation for an arbitrary operator, we can
apply it to the angular momentum operators, the direction operators, the collision operator and the
rotation operators. We note that $\mathbf{O}_{R}$ contains complex terms and, thus, we have to check
if the resulting representation matrices corresponding to the above list of operators are real.

When computing the matrix representations with respect to the real spherical harmonics, it is
important to note that $m', m \geq 0$, hence $\delta_{-m'(m+1)} = \delta_{m'(-m-1)} = 0$ and
$\delta_{-m'(m - 1)} = \delta_{m'(-m + 1)} = \delta_{m'0}\delta_{m1} + \delta_{m'1}\delta_{m0}$.
Moreover, $m'=0$ and $s'=1$ (or $m=0$ and $s=1$) is excluded\footref{fn:no-m-zero-s-one}, which we
use to remove terms from the resulting expressions. Generally, we remove terms with combinations of
Kronecker deltas whose evaluation leads to values of $l, m$ and $s$ which are not allowed.

We begin with the representation matrices of the angular momentum operator. We remind ourselves that
we have to include a factor $\complexi$, because
$\bm{\Omega}^{a}_{R} = \complexi \mathbfss{S}^{\bm{\dagger}}\bm{\Omega}^{a}\mathbfss{S}$. Using
eq.~\eqref{eq:real-representation-matrix-arbitrary-operator} yields
\begin{equation}
  \label{eq:real-omega-matrices}
  \begin{split}
 & (\bm{\Omega}^{x}_{R})_{i(l',m',s')j(l,m,s)}                                                                                                                                                                             \\
 & \qquad = m \delta_{l'l}\delta_{m'm}
   \left(\frac{\delta_{s'0}\delta_{s1}}{\sqrt{1 + \delta_{m'0}}} - \frac{\delta_{s'1}\delta_{s0}}{\sqrt{1 + \delta_{m0}}}\right) 
    = \bm{\Omega}^{x}_{i(l',m')j(l,m)}\left(\frac{\delta_{s'0}\delta_{s1}}{\sqrt{1 + \delta_{m'0}}} - \frac{\delta_{s'1}\delta_{s0}}{\sqrt{1 + \delta_{m0}}}\right)                                                        \\
 & (\bm{\Omega}^{y}_{R})_{i(l',m',s')j(l,m,s)}                                                                                                                                                                             \\
 & \qquad = \frac{\delta_{l'l}}{2}\Biggl[\Biggr.\frac{\delta_{s'0}\delta_{s1}}{\sqrt{1 + \delta_{m'0}}}
   \left( \sqrt{(l + m +1)(l-m)} \delta_{m'(m+1)} + \sqrt{(l + m)(l - m + 1)}\delta_{m'(m-1)}  +  \sqrt{l(l+1)} \delta_{m'0}\delta_{m1}  \right)                                                                           \\
 & \qquad  \phantom{=} \qquad \, \, \Biggl. {} - \frac{\delta_{s'1}\delta_{s0}}{\sqrt{1 + \delta_{m0}}}
   \left( \sqrt{(l + m +1)(l-m)} \delta_{m'(m+1)} + \sqrt{(l + m)(l - m + 1)}\delta_{m'(m-1)}  +  \sqrt{l(l+1)}\delta_{m'1}\delta_{m0}  \right)\Biggr]                                                                     \\
 & \qquad = \frac{\delta_{s'0}\delta_{s1}}{\sqrt{1 + \delta_{m'0}}} \left( \bm{\Omega}^{y}_{i(l', m')j(l, m)}  +  \frac{\delta_{l'l}}{2}\sqrt{l(l+1)} \delta_{m'0}\delta_{m1}  \right) 
   - \frac{\delta_{s'1}\delta_{s0}}{\sqrt{1 + \delta_{m0}}} \left(\bm{\Omega}^{y}_{i(l',m')j(l,m)}  +  \frac{\delta_{l'l}}{2}\sqrt{l(l+1)}\delta_{m'1}\delta_{m0}  \right)                                                 \\
 & (\bm{\Omega}^{z}_{R})_{i(l',m',s')j(l,m,s)}                                                                                                                                                                             \\
 & \qquad = \frac{\delta_{l'l}}{2}\Biggl[\Biggr.\frac{\delta_{s'0}\delta_{s0}}{\sqrt{(1 + \delta_{m'0})(1 + \delta_{m0})}}
                                                 \left( \sqrt{(l + m +1)(l-m)} \delta_{m'(m+1)} - \sqrt{(l + m)(l - m + 1)}\delta_{m'(m-1)}  -  \sqrt{l(l+1)} (\delta_{m'0}\delta_{m1} - \delta_{m'1}\delta_{m0})  \right) \\
 & \qquad \phantom{=} \qquad \, \, \Biggl. {} + \delta_{s'1}\delta_{s1}
   \left( \sqrt{(l + m +1)(l-m)} \delta_{m'(m+1)} - \sqrt{(l + m)(l - m + 1)}\delta_{m'(m-1)}  \right)\Biggr]                                                                                                              \\
 & \qquad = \frac{\delta_{s'0}\delta_{s0}}{\sqrt{(1 + \delta_{m'0})(1 + \delta_{m0})}}
                                                 \left( \complexi \bm{\Omega}^{z}_{i(l',m')j(l,m)}  -  \frac{\delta_{l'l}}{2}\sqrt{l(l+1)} (\delta_{m'0}\delta_{m1} - \delta_{m'1}\delta_{m0})  \right)
  + \delta_{s'1}\delta_{s1} \complexi \bm{\Omega}^{z}_{i(l',m')j(l,m)} 
  \end{split}
\end{equation}

We proceed with the matrix representations of the direction operators $\operator{A}^{a}$ operators.
We find that
\begin{equation}
  \label{eq:real-A-matrices}
  \begin{split}
 & (\mathbfss{A}^{x}_{R})_{i(l',m',s')j(l,m,s)}                                                                                                                                                                                                                                    \\
 & \qquad  = \delta_{m'm}\delta_{s's} \left(\sqrt{\frac{(l + m + 1)(l - m +1)}{(2l + 3)(2l + 1)}} \delta_{l'(l+1)} + \sqrt{\frac{(l+m)(l-m)}{(2l + 1)(2l - 1)}} \delta_{l'(l-1)}\right)
    = \delta_{s's} \mathbfss{A}^{x}_{i(l',m')j(l,m)}                                                                                                                                                                                                                               \\
 & (\mathbfss{A}^{y}_{R})_{i(l',m',s')j(l,m,s)}                                                                                                                                                                                                                                    \\
 & \qquad =  \frac{\delta_{s'0}\delta_{s0}}{\sqrt{(1 + \delta_{m'0})(1 + \delta_{m0})}} \left( (\mathbfss{A}^{y})_{i(l',m')j(l,m)} + \frac{1}{2} \left(a^{-m+1}_{l+1}\delta_{l'(l+1)} - a^{m}_{l}\delta_{l'(l-1)}\right)(\delta_{m'0}\delta_{m1} - \delta_{m'1}\delta_{m0})\right)  + \delta_{s'1}\delta_{s1}(\mathbfss{A}^{y})_{i(l',m')j(l,m)}                                        \\
 & (\mathbfss{A}^{z}_{R})_{i(l',m',s')j(l,m,s)}                                                                                                                                                                                                                                    \\
 & \qquad =  \frac{\delta_{s'0}\delta_{s1}}{\sqrt{1 + \delta_{m'0}}} \left( -\complexi (\mathbfss{A}^{z})_{i(l',m')j(l,m)} + \frac{1}{2} \left(a^{-m+1}_{l+1}\delta_{l'(l+1)} - a^{m}_{l}\delta_{l'(l-1)}\right)\delta_{m'0}\delta_{m1} \right)        \\
 & \qquad \phantom{=} {} - \frac{\delta_{s'1}\delta_{s0}}{\sqrt{1 + \delta_{m0}}}\left(-\complexi (\mathbfss{A}^{z})_{i(l',m')j(l,m)}  +  \frac{1}{2} \left(a^{-m+1}_{l+1}\delta_{l'(l+1)} - a^{m}_{l}\delta_{l'(l-1)}\right)\delta_{m'1}\delta_{m0}\right)
  \end{split}
\end{equation}
We point out that $\mathbfss{A}^{x} = \mathbfss{A}^{x}_{R}$.

The same is true for the real representation matrix of the collision operator $\operator{C}$, which
is
\begin{equation}
  \label{eq:real-C-matrix}
  (\mathbfss{C}_{R})_{i(l'm's')j(l,m,s)} = \frac{l(l + 1)}{2}\delta_{l'l}\delta_{m'm}\delta_{s's} \,.
\end{equation}

It is left to compute the real representation matrices of the used rotation operators, i.e.
$e^{\complexi \frac{\upi}{2} \operator{L}_{x}}$ and $e^{\complexi \frac{\upi}{2} \operator{L}_{z}}$.
They are
\begin{equation}
  \label{eq:real-U-matrices}
  \begin{split}
    (\mathbfss{U}_{R}(-\upi/2\mathbfit{e}_{x}))_{i(l',m',s')j(l,m,s)} & =  e^{\frac{\upi}{2}\bm{\Omega}^{x}_{R}}             \\
                                                                     & = \delta_{l'l}\delta_{m'm}\left[\frac{\delta_{s'0}\delta_{s0}}{\sqrt{(1 + \delta_{m'0}) (1 + \delta_{m0})}} \left(\cos\left(\frac{\upi}{2}m\right)  + 1 \delta_{m'0}\delta_{m0}\right)
        + \frac{\delta_{s'1}\delta_{s0}}{\sqrt{1 + \delta_{m'0}}} \sin\left(\frac{\upi}{2} m\right) \right. \\
                                                                     & \phantom{= \delta_{l'l}\delta_{m'm}} \left. \, \, \, {} - \frac{\delta_{s'0}\delta_{s1}}{\sqrt{1 + \delta_{m0}}} \sin\left(\frac{\upi}{2} m\right)
                                                                       + \delta_{s'1}\delta_{s1} \cos\left(\frac{\upi}{2} m\right)\right] \\
    (\mathbfss{U}_{R}(-\upi/2\mathbfit{e}_{z}))_{i(l',m',s')j(l,m,s)} & =  e^{\frac{\upi}{2}\bm{\Omega}^{z}_{R}}\\
                                                                     & = \Biggl[\frac{\delta_{s'0}\delta_{s0}}{\sqrt{(1 + \delta_{m'0}) (1 + \delta_{m0})}}\left((\mathbfss{U}(-\upi/2\mathbfit{e}_{z})_{i(l',m')j(l,m)} + (-1)^{m} \mathbfss{U}(-\upi/2\mathbfit{e}_{z})_{i(l',m')j(l,-m)}) \right)\Biggr. \\
    &\phantom{=} \Biggl. \, \, \, {} + \delta_{s'1}\delta_{s1}\left((\mathbfss{U}(-\upi/2\mathbfit{e}_{z})_{i(l',m')j(l,m)} - (-1)^{m} \mathbfss{U}(-\upi/2\mathbfit{e}_{z})_{i(l',m')j(l,-m)}) \right) \Biggr]
  \end{split}
\end{equation}
For the last equation we assumed that $\mathbfss{U}_{R}(-\upi/2\mathbfit{e}_{z})$ is real, which is
of course against the idea to prove that it is real. Since $\mathbfss{U}(-\upi/2\mathbfit{e}_{z})$ is
real as well, we concluded that the two terms in middle of
eq.~\eqref{eq:real-representation-matrix-arbitrary-operator} must be zero for
$\mathbfss{U}_{R}(-\upi/2\mathbfit{e}_{z})$. This conclusion lead us to conjecture that
$(-1)^{m'}\mathbfss{U}(\upi/2\mathbfit{e}_{z})_{i(l',-m')j(l,m)} = (-1)^{m}
\mathbfss{U}(-\upi/2\mathbfit{e}_{z})_{i(l',m')j(l,-m)}$ and that
$\mathbfss{U}(-\upi/2\mathbfit{e}_{z})_{i(l',m')j(l,m)} = (-1)^{m' + m}
\mathbfss{U}(-\upi/2\mathbfit{e}_{z})_{i(l',m')j(l,-m)}$. These conjectures give us the above result
and numerical experiments convinced us that these conjectures are true.

\bsp	
\label{lastpage}
\end{document}